\documentclass[iop]{emulateapj}
\usepackage{color}

\pdfoutput=1

\newcommand{\sersic}{S\'{e}rsic}

\slugcomment{Merging Red Quasar Hosts at $z\sim 2$}

\shorttitle{Merging Red Quasar Hosts at $z\sim 2$}
\shortauthors{Glikman et al.}

\begin{document}


\title{Major Mergers Host the Most Luminous Red Quasars at $z\sim 2$: A {\em Hubble Space Telescope WFC3/IR} Study.\footnote{Based on GO observations made with the NASA/ESA Hubble Space Telescope, as well as from the Data Archive at the Space Telescope Science Institute, which is operated by the Association of Universities for Research in Astronomy, Inc., under NASA contract NAS 5-26555. These observations are associated with program \# 12942.}}

\author{Eilat Glikman\altaffilmark{1}, Brooke Simmons\altaffilmark{2}, Madeline Mailly\altaffilmark{1}, Kevin Schawinski\altaffilmark{3}, C. M. Urry\altaffilmark{4}, M. Lacy\altaffilmark{5}}


\altaffiltext{1}{Middlebury College}
\altaffiltext{2}{Oxford}
\altaffiltext{3}{Institute for Astronomy, Department of Physics, ETH Zurich, Wolfgang-Pauli-Strasse 27, CH-8093 Zurich, Switzerland}
\altaffiltext{4}{Yale University}
\altaffiltext{5}{NRAO Charlottesville}


\begin{abstract}
We used the {\em Hubble Space Telescope WFC3} near-infrared camera to image the host galaxies of a sample of eleven luminous, dust-reddened quasars at $z\sim 2$ -- the peak epoch of black hole growth and star formation in the Universe -- to test the merger-driven picture for the co-evolution of galaxies and their nuclear black holes.
The red quasars come from the FIRST+2MASS red quasar survey and a newer, deeper, UKIDSS+FIRST sample. 
These dust-reddened quasars are the most intrinsically luminous quasars in the Universe at all redshifts, and  may represent the dust-clearing transitional phase in the merger-driven black hole growth scenario.
Probing the host galaxies in rest-frame visible light, the HST images reveal that 8/10 of these quasars have actively merging hosts, while one source is reddened by an intervening lower redshift galaxy along the line-of-sight. We study the morphological properties of the quasar hosts using parametric \sersic\ fits as well as the non-parametric estimators (Gini coefficient, $M_{20}$ and asymmetry).  Their properties are heterogeneous but broadly consistent with the most extreme morphologies of local merging systems such as Ultraluminous Infrared galaxies.  The red quasars have a luminosity range of $\log(L_{\rm bol}) = 47.8 - 48.3$ (erg s$^{-1}$) and the merger fraction of their AGN hosts is consistent with merger-driven models of luminous AGN activity at $z=2$, which supports the picture in which luminous quasars and galaxies co-evolve through major mergers that trigger both star formation and black hole growth. \\
\end{abstract}




\section{Introduction}
Most galaxies in our Universe have supermassive black holes (SMBH) at their centers, which are thought to have grown in earlier epochs when galaxies had a higher gas content. Quasars --- highly luminous evidence of rapidly accreting black holes --- provide insight into this important stage of galaxy evolution. In particular, dust-reddened quasars, such as those investigated in this project, are the most intrinsically luminous steadily-emitting objects in the Universe and may represent an intermediate stage between galaxy mergers and luminous blue quasars, which eventually become quiescent SMBHs. 

Models of gas rich galaxy mergers have been proposed to explain the observed link between the growth of supermassive black holes (SMBHs) and their host galaxies by an evolutionary scenario in which the growing black hole moves from a heavily-enshrouded high-accretion phase as in observations of some local ultraluminous infrared galaxies (ULIRGs) through a brief ``blow-out'' phase where winds and outflows clear the obscuring dust, to an unobscured, blue quasar which later becomes a quiescent black hole \citep{Sanders88,Hopkins05c,Hopkins08}.  In this scenario, the so-called ``blow-out'' phase is expected to appear as a reddened Type I quasar (i.e., showing broad emission lines in its spectrum) with high Eddington ratios and strong outflows.  Systems in this phase are also expected to still show signs of the earlier, or possibly still ongoing, merger in the form of tidal tails and disturbed morphologies.  

Dust-reddened quasars (or ``red quasars''), found predominantly in samples of matched radio and near-infrared sources, satisfy the conditions for this blow-out phase.  The largest sample of red quasars  comes from the FIRST-2MASS sample \citep[called F2M hereafter;][]{Glikman04,Glikman07,Urrutia09,Glikman12}.  More recently, \citet{Glikman13} matched the FIRST radio survey to the United Kingdom Infrared Telescope (UKIRT) Infrared Deep Sky Survey \citep[UKIDSS;][]{Lawrence07,Warren07} over a pilot area of 190 deg$^2$, identifying twelve new, fainter red quasars, three of them above $z\sim 2$ (prefixed with UKFS). In total, our group has identified over 135 unique red quasars in the redshift range $0.1<z\lesssim3$.  

\citet{Glikman12}  showed that, when corrected for extinction, F2M red quasars are {\em more luminous} than blue quasars at every redshift.  
In addition, the fraction of red quasars {\em increases} with intrinsic luminosity, in agreement with the blow-out model.  
Other surveys of obscured AGN \citep[e.g.,][]{Assef12} find that the obscured fraction rises with {\em decreasing} luminosity consistent with a higher covering fraction from a dusty, axisymmetric structure and consistent with a unified model of AGN \citep{Urry95}.  At higher luminosities, this so-called ``receding torus'' is expected to vanish due to dust sublimation \citep{Lawrence91}.  Therefore, the reddening seen for red quasars is probably due not to a circumnuclear torus but to dust distributed throughout the host galaxy that is created during a merger-induced starburst. 
We are able to disentangle the two effects with the F2M sample because F2M red quasars are selected to have broad emission lines in their spectra indicating a viewing angle that avoids toroidal extinction.

A subsample of  13 F2M quasars at $z\sim 0.7$ were imaged with the {\em Hubble Space Telescope} ({\em HST}) Advanced Camera for Surveys (ACS) by \citet{Urrutia08}.  The images showed that  F2M quasars are hosted by an unusually high fraction of mergers and/or interacting systems.  While studies of the host galaxies of unreddened quasars find a merger fraction of $\lesssim 30\%$ \citep{Dunlop03,Floyd04}, the red quasar merger fraction is 85\%. In addition, {\em Spitzer} observations of the same 13 red quasars reveal that most of these objects are accreting at very high Eddington rates \citep{Urrutia12}. These results support the picture that red quasars are an early dust-enshrouded phase of quasar-galaxy co-evolution. 

Despite the successes of these aforementioned merger-driven models, recent observations suggest a more complicated picture for AGN generally. \citet{Schawinski11}, \citet{Simmons12} and \citet{Kocevski12} showed that most  moderate-luminosity X-ray-selected AGN ($10^{42} {\rm~erg~s}^{-1} < L_X < 10^{44} {\rm~erg~s}^{-1}$) at $1.5 < z < 3$ reside in undisturbed, disk-dominated galaxies. 
\citet{Schawinski12} showed that this is also the case for $L_{\rm bol} \sim 10^{45}  {\rm~erg~s}^{-1}$ heavily obscured quasars at $z\sim2$.

Although some theoretical models imply that the presence of a disk does not in-itself eliminate the possibility of a merger, as disks can survive mergers or can be rebuilt quickly \citep{Puech12}, many models allow for stochastic accretion to dominate at low luminosities (and low black hole masses and/or low accretion rate) while mergers drive fueling at high luminosities \citep[e.g.,][]{Hopkins06d}.  This dependence on luminosity is supported by a recent meta-analysis of mergers in AGN hosts by \citet{Treister12} in which the merger fraction rises monotonically over three orders of magnitude in bolometric luminosity, with the highest merger rate (85\%) at the highest luminosity bin ($L\sim 10^{46}$ erg s$^{-1}$) represented by the \citet{Urrutia08} {\em HST} imaging study of red quasars. 

However compelling, this picture has not been tested at $z\sim 2$, the epoch of peak quasar activity, where most SMBH growth is believed to occur. 
A handful of luminous blue quasars have been studied at $z \sim 2$ with the Near-Infrared Camera and Multi-Object Spectrometer (NICMOS) on {\em HST} \citep{Kukula01,Ridgway01}.  Those studies were challenged by the removal of the bright point source to study their hosts' morphologies; only host luminosities were reported. This may be because obvious signs of mergers have already faded in these mature systems.  In the present study,  we use {\em HST} observations with the near-infrared detector of the Wide Field Camera 3 (WFC3/IR) to examine red quasars in the rest-frame optical at $z\sim2$, to probe this critical stage of black hole growth and galaxy evolution.

The paper is organized as follows.  Section \ref{sec:selection} describes the parent samples and selection of $z\sim2$ red quasars observed with {\em HST} WFC3/IR.  Section \ref{sec:obs} describes the {\em HST} observations and the data reduction procedure.  Section \ref{sec:astro_phot} provides photometric and source count statistics for the observed fields and Section \ref{sec:galfit} describes multi-component parametric fitting to separate the quasars from their host galaxies.  We discuss the derived properties of the quasars and their host galaxies in Section \ref{sec:results} followed by a discussion of the individual quasars in Section \ref{sec:indiv_qsos}.  We address the implications in Section \ref{sec:disc} and summarize our findings in Section \ref{sec:summ}.

Throughout this work we quote magnitudes on the AB system, unless explicitly stated otherwise.  When computing luminosities and any other cosmology-dependent quantities, we use the $\Lambda$CDM concordance cosmology: $H_0=70$ km s$^{-1}$ Mpc$^{-1}$, $\Omega_M=0.30$, and $\Omega_\Lambda=0.70$.

\section{Quasar Sample Selection}\label{sec:selection}

\begin{figure}
\plotone{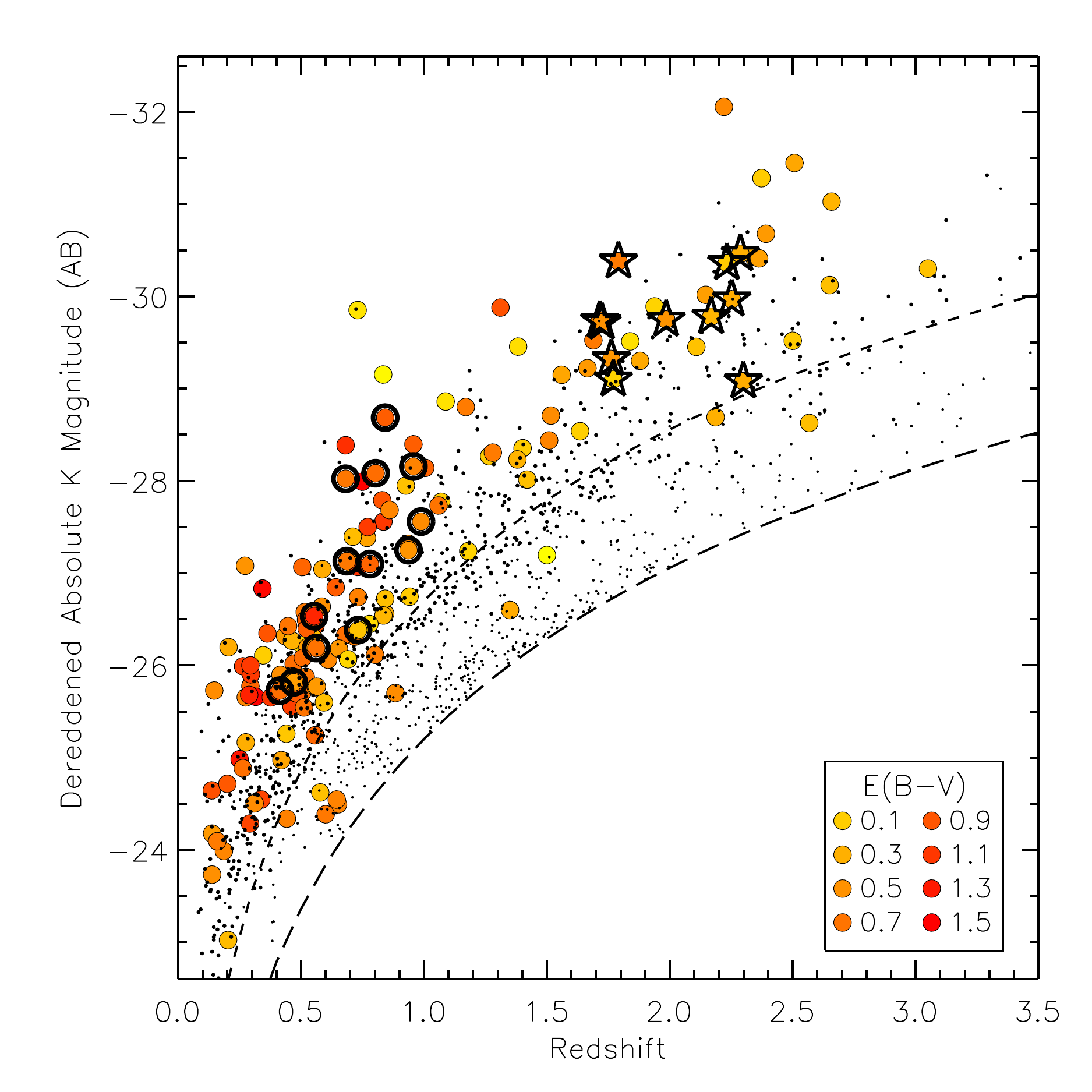}
\caption{Dereddened K-band absolute magnitude of F2M and UKFS quasars vs.~redshift. 
The respective $K$-band limits of the F2M survey ($K_{AB}<17.4$) and UKFS ($K_{AB}<18.9$) are indicated with dashed and long-dashed line.  
Colored circles correspond to the amount of reddening, as defined in the legend. The small black dots are blue, optically-selected quasars from the FIRST bright quasar survey \citep[FBQS;][]{Gregg96} and radio-detected quasars from a deeper SDSS+UKIDSS catalog \citep{Peth11} for which no reddening is assumed. The HST-imaged objects from \citet{Urrutia08} are marked with thick black circles and the sources studied in this work are marked with stars. }\label{fig:absKz}
\end{figure}

The F2M red quasars at $z \sim 2$ present an ideal sample to test merger-driven co-evolution in the highest luminosity regime. 
Figure \ref{fig:absKz} shows the de-reddened $K$-band absolute magnitude for red quasars (colored circles) vs. redshift.  For comparison, we plot with black dots unreddened blue quasars with FIRST detections from the FIRST Bright Quasar Survey \citep[FBQS;][]{Gregg96} and SDSS+UKIDSS \citep{Peth11}.  The quasars marked with thick circles are the thirteen objects previously studied with ACS on {\em HST} by \citet{Urrutia08}.  The sources marked with stars are the eleven quasars studied in this work. The sample spans about two magnitudes in intrinsic luminosity and the redshift range of $1.7 < z < 2.3$.

As Figure \ref{fig:absKz} shows, F2M quasars are among the most luminous quasars in the Universe after correcting for extinction.  At the redshift range of our sample, the WFC3/IR $H-$band is comparable to the ACS $I-$band, which makes this study a high-redshift analog of the $z\sim0.7$ sample studied by \citet{Urrutia08}.  
Although the $z\sim 2$ sample is more luminous by $\sim 2-3$ magnitudes, this increase in luminosity is consistent with the evolution of $L_*$ in the quasar luminosity function \citep[QLF;][]{Croom09} and thus means we are sampling the same portion of the QLF in both studies.

We selected for {\em HST} imaging all ten F2M red quasars from \citet{Glikman12} with $1.7 < z < 2.3$ that had a near-infrared spectrum.  
To extend our sample to fainter magnitudes, we added an eleventh quasar from the UKFS sample \citep{Glikman13} which also possessed a near-infrared spectrum. Table \ref{tab:qso_sample} lists our targets, their near infrared magnitudes, their $K$-band absorption and redshift. Note that these quasars experience $K$-band absorption ranging between $A_K = 0.3$ and $A_K = 1.2$ magnitudes. 

\begin{figure*}
\plottwo{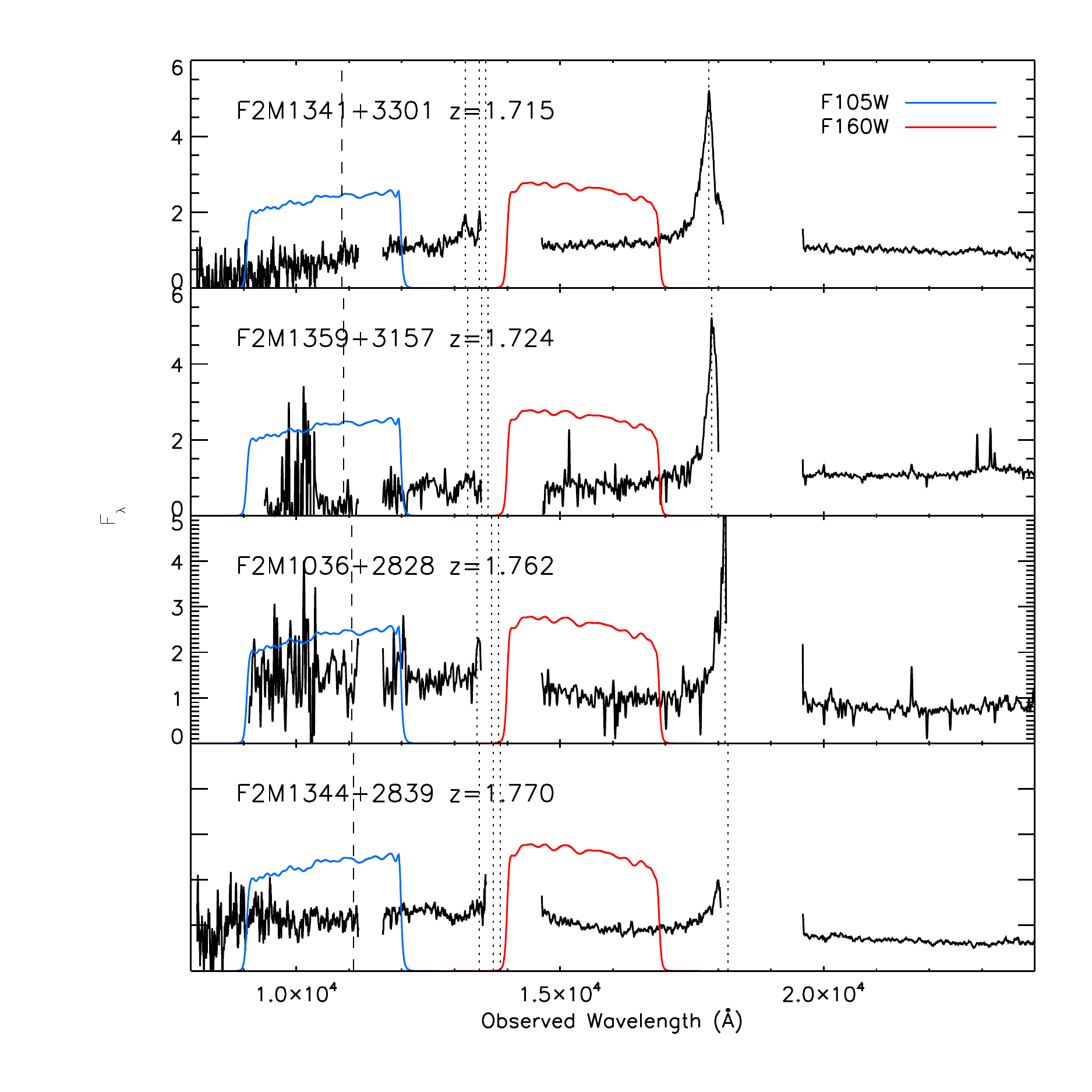}{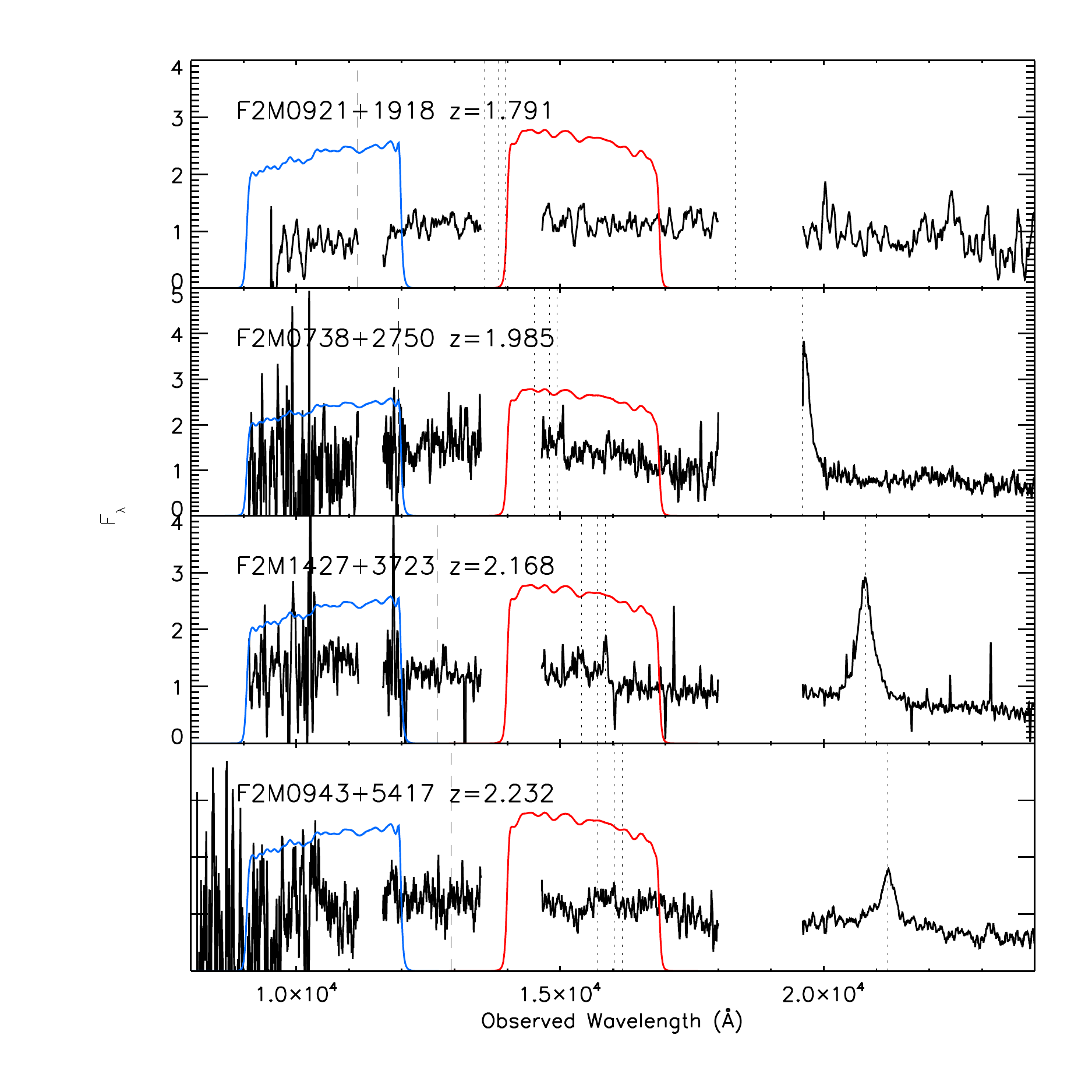}
\caption{Near-infrared spectra of the eight targets imaged with F160W and F105W filters. The WFC3 filter curves are shown in each panel in blue and red, respectively.  
The locations of H$\alpha$ and H$\beta$+[\ion{O}{3}] are indicated by vertical dotted lines and the host galaxy's 4000\AA\ break is shown with a dashed line. Note that in a few cases the Balmer lines are shifted into the atmospheric absorption bands and are not seen.  For these objects we determined the redshifts from optical spectra, which were presented in Figure 6 of \citet{Glikman12}.  The objects' redshifts ensure that the strong emission from H$\alpha$ does not enter the F160W bandpass, minimizing the quasar/galaxy ratio and enabling more accurate PSF subtraction.}\label{fig:spectra}
\hspace{0.5in}
\end{figure*}

\begin{figure}
\plotone{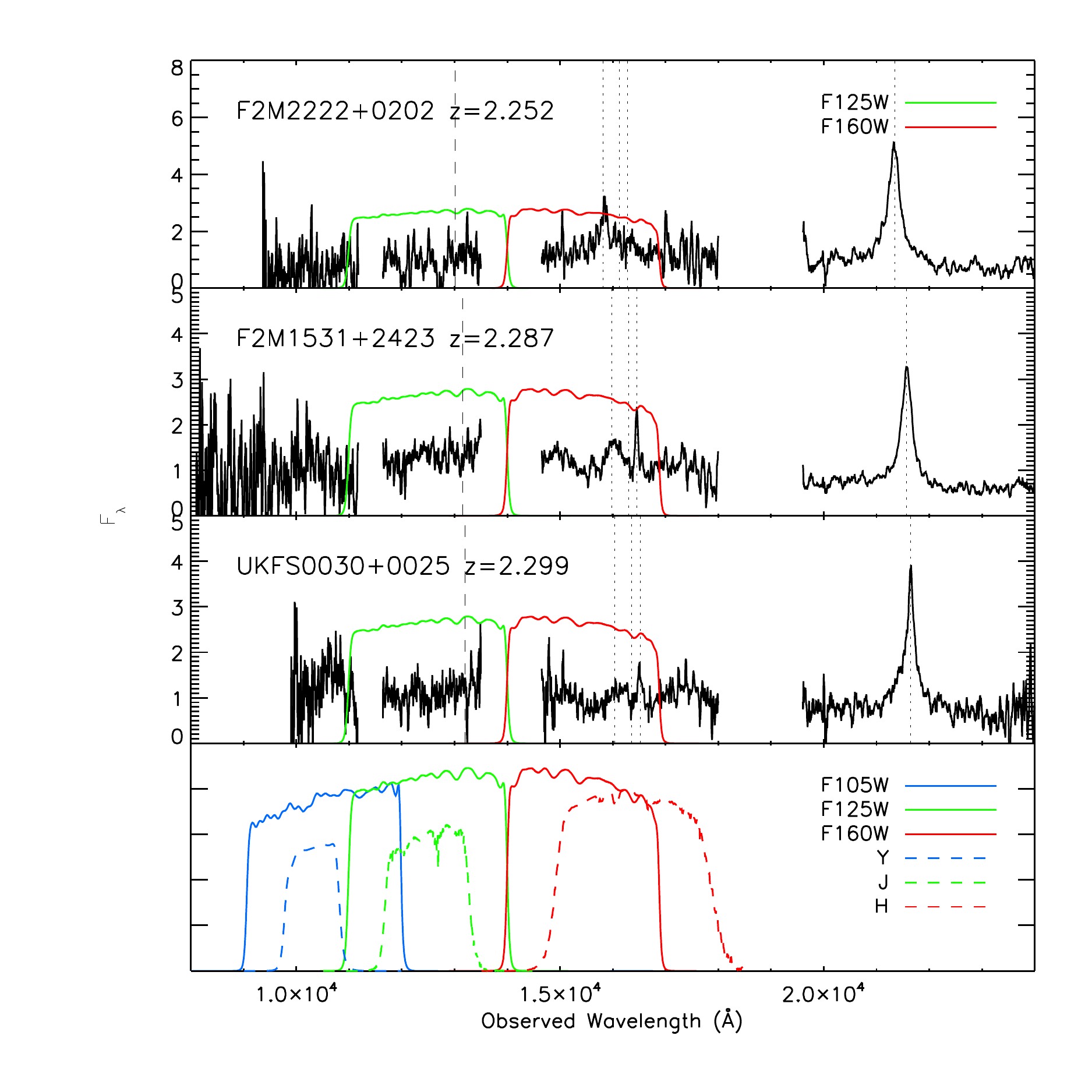}
\caption{Same as \ref{fig:spectra}, except for the three highest redshift sources in our sample that were imaged with F160W and F125W filters.   The bottom panel shows the transmission curves for the three WFC3/IR filters used in this study. For comparison we show the UKIDSS $Y$, $J$, and $H$ filters with dashed lines. }\label{fig:spectra2}
\hspace{0.5in}
\end{figure}

\section{Observations and Image Reduction}\label{sec:obs}

The {\em Hubble Space Telescope} WFC3/IR camera is sensitive to near-infrared wavelengths from 800 to 1700 nm and offers 15 different filters of narrow, medium and wide wavelength transmission range.  The pixel scale is 0.12825\arcsec/pixel and the field of view of the IR detector is $136\arcsec \times 123\arcsec$. For this study, we used the F160W, F125W, and F105W filters, whose effective wavelengths are 1536.9 nm, 1248.6 nm, and 1055.2 nm, respectively. Images of each quasar were obtained in two filters -- the F160W filter and either the F125W filter (for the three quasars with the highest redshifts) or the F105W filter. 

Figure \ref{fig:spectra} shows the near-infrared spectra of the eight quasars imaged with F160W and F105W and the transmission curves of those filters.  Figure \ref{fig:spectra2} shows the same but for the three highest redshift sources, whose blue images were taken with the F125W filter. Note that in no case is the broad WFC3 F160W filter contaminated by the strong H$\alpha$ emission line.  The avoidance of strong emission lines and the large amounts of extinction in these quasars minimizes contamination from the quasar and allows for better sensitivity to low surface brightness features in the host galaxies. We also image our sources in a bluer filter to sample the host galaxy light below the 4000\AA\ break (marked with a vertical dashed line in the Figure), enabling us to study the host galaxies' star formation properties.  Table \ref{tab:qso_sample} lists the WFC3/IR filters used.

The bottom panel of Figure \ref{fig:spectra2} shows the transmission curve of the three filters used in this study (solid lines) compared with the $Y$, $J$ and $K_s$ bands used in ground based imaging, such as UKIDSS \citep[dashed lines;][]{Hewett06}.

We observed most of the quasars with both filters over a single orbit, reaching a 3$\sigma$ surface brightness of $\sim$24 AB magnitudes arcsec$^{-2}$ per pixel-area in all bands.  We report this value in a columns (11) and (14) of Table  \ref{tab:qso_sample} (see \S \ref{sec:results} for details on the surface brightness depth of the images).   

The WFC3/IR detector is capable of non-destructive readouts (NDRs) during an exposure and has several options for NDR sequencing depending on the dynamic range desired in an image.  We observed our sources in MULTIACCUM mode using the STEP sampling which is a log-linear non-destructive readout mode that prevents saturation of bright stars and allows a broad dynamic range in a single exposure. Depending on the total exposure time, we used the STEP25, STEP50, or STEP100 sequences.  The observations were done in a 4-point box dither pattern, which helps improve the resolution of the final reduced images and enables sub-pixel sampling of the point spread function (PSF) which we drizzle to a final pixel scale 0f 0.06\arcsec/pixel.  The total exposure times of the reduced images are listed in columns (10) and (13) of Table  \ref{tab:qso_sample} for the blue and red filters, respectively.

\begin{figure*}
\epsscale{0.9}
\plotone{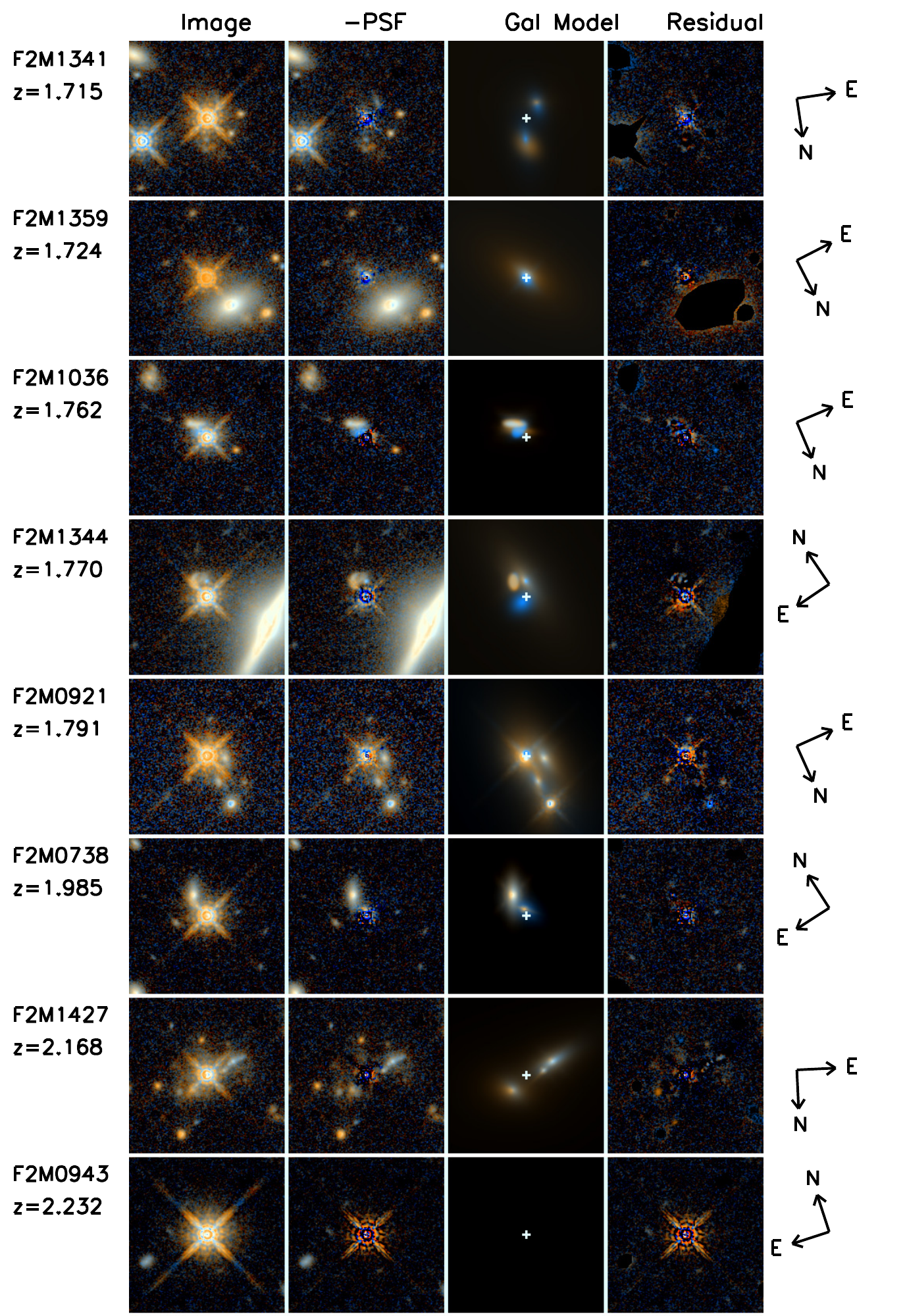}
\caption{Two color HST images of the 8 lower-redshift quasars studied in this paper imaged with F105W and F160W. Each row represents a separate object.  The first column is the original image shown at a scale of $8\arcsec \times 8\arcsec$.  The second column shows the residual image after subtracting only the point source component.  The third column shows the model for all but the point-source component; the blank frame is a source to which no host component could be fit. The final panel shows the full residual including masked regions and is indicative of the overall goodness of fit. 
Evidence of mergers and disrupted host galaxies is seen in most the sources. We apply the Red-Green-Blue color-combining algorithm of \citet{Lupton04} to our images, and we average the count rate from the F105W and F160W images to produce the green frame. }\label{fig:hst_images}
\hspace{0.5in}
\end{figure*}

\begin{figure*}
\plotone{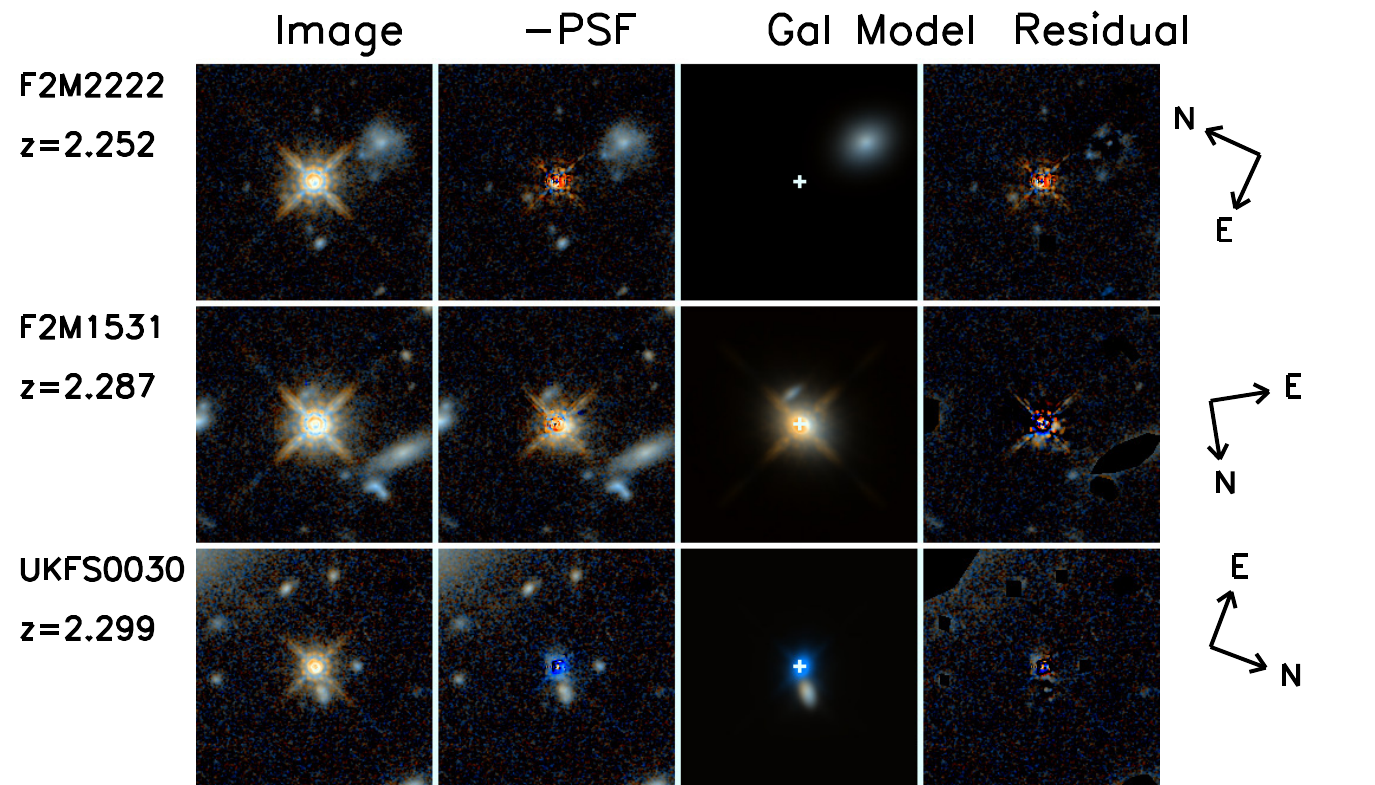}
\caption{Same as \ref{fig:hst_images} except here we show the three highest-redshift quasars imaged with F125W and F160W. }\label{fig:hst_images2}
\hspace{0.5in}
\end{figure*}

Figures \ref{fig:hst_images} and \ref{fig:hst_images2} show the {\em HST} images of the eleven quasars.  Visual inspection suggests that most have a nearby companion and/or a disrupted host. Although visual inspection has been used in previous works to identify mergers in high redshift systems \citep{Kocevski12,Schawinski12} we also performed careful two-dimensional modeling of the point source plus host galaxy (presented in \S \ref{sec:galfit}).  However, to better define the parameters of our images we first made astrometric and photometric measurements of sources in each field and used them to determine the significance of the nearby sources as true companions.

\section{Astrometry, Photometry and Source Distribution}\label{sec:astro_phot}

Although the relative astrometry of the WFC3/IR images is highly accurate (10 mas, according to the WFC3 Data Handbook), the absolute astrometry of the images can be offset by as much as 1.5\arcsec\ from the absolute astrometric grid\footnote{\tt http://www.stsci.edu/hst/wfc3/documents/handbooks/currentDHB/wfc3\_Ch74.html}.  To determine the offsets and correct for them, we extracted source catalogs from each reduced science frame plus its associated weight image using the SExtractor software package \citep{Bertin96} with a 5$\sigma$ detection limit.  We used these catalogs to match to SDSS for absolute astrometric correction and to UKIDSS to test and correct for deviations from the photometric zero-point provided by the WFC3 Guide\footnote{\tt http://www.stsci.edu/hst/wfc3/phot\_zp\_lbn}.  

\begin{figure}
\plotone{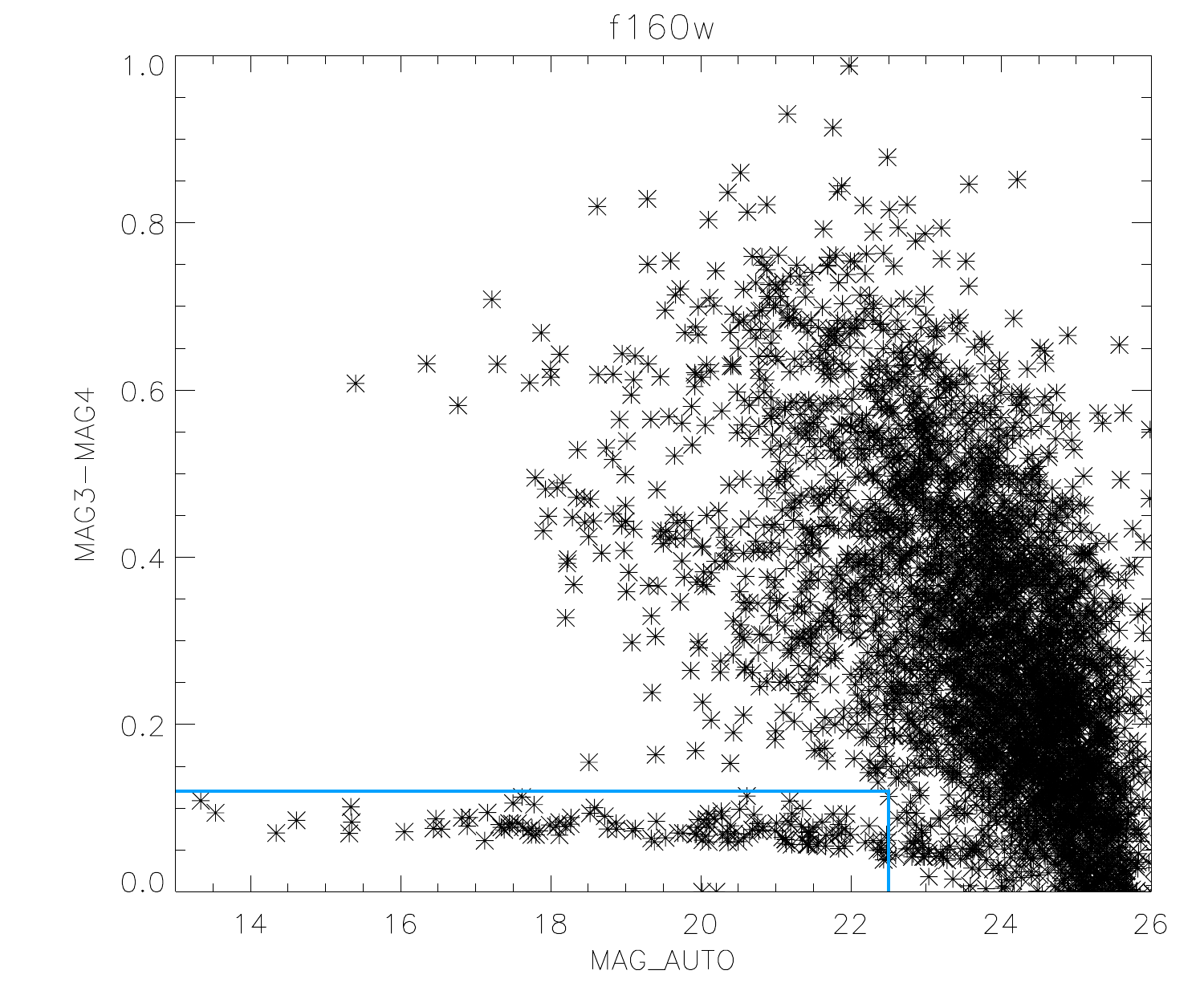}
\caption{Differential aperture photometry for all the sources catalogued in our F160W fields showing the clear separation of point-like objects which lie in a tight locus (indicated by the blue region) from the cloud of extended sources.  }\label{fig:stars}
\hspace{0.5in}
\end{figure}

We identified point sources by performing differential photometry on our catalogued sources and plotting $\Delta$mag in two apertures (in this case 12 and 20 pixels, or 0.72\arcsec and 1.2\arcsec) versus the MAG\_AUTO which is a better estimate the source's total flux than the aperture magnitudes.  An example of this analysis for the F160W filter is shown in Figure \ref{fig:stars}.
Since SDSS matches will necessarily be brighter than most of the objects in the {\em HST} image, we restrict SExtractor to  5$\sigma$ sources brighter than 22.5 magnitudes (AB)  whose differential aperture photometry lies along a constant locus, separate from galaxies.  In each field we find between 6 and 20 stars that we analyze on a field-by-field basis to determine astrometric offsets.  

Using astrometrically-corrected images, we compared the position of the FIRST radio image with the {\em HST} images.  Similar to the findings of \citet{Urrutia08}, we find that the radio peak overlaps the peak of the WFC3 images.  Since the angular resolution of FIRST images is 5\arcsec\ (and pixel size of 1\farcs8), we could not compare structural details of the radio source with any structure in our {\em HST} images.  

We also matched our catalogued sources to the UKIDSS DR9 LAS survey (using the multiple cone search tool in TOPCAT).  Only five our fields have UKIDSS coverage (F2M1341, F2M1359, F2M1344, F2M0738, UKFS0030).  The first four are imaged with F105W while the only F125W fields covered by UKIDSS is UKFS0030.   To perform a photometric comparison between our WFC3 filters and traditional near-infrared filters such as those provided by UKIDSS, we shift the AB magnitudes of the point sources in our {\em HST} images to their Vega counterparts using the AB to Vega corrections given by \citet{Hewett06}.
 We compare the F105W, F125W and F160W magnitudes to the UKIDSS $Y$, $J$, and $H$ bands, respectively.  We find that the two magnitudes are well fit by a line of slope unity.  The intercept of the line fit represents an offset between the two magnitude systems (i.e., add $0.58$, $0.49$, $0.80$ magnitudes, respectively) which encompass differences in the filter transmission curves between the WFC3/IR broad band filters and the UKIDSS filters, as is evident in the bottom panel of Figure \ref{fig:spectra2}.  
We report magnitudes in the rest of the paper using the WFC3/IR zeropoints on the AB system.  The shifts are provided above to enable us to make quantitative statements in either system.

\begin{figure*}
\plottwo{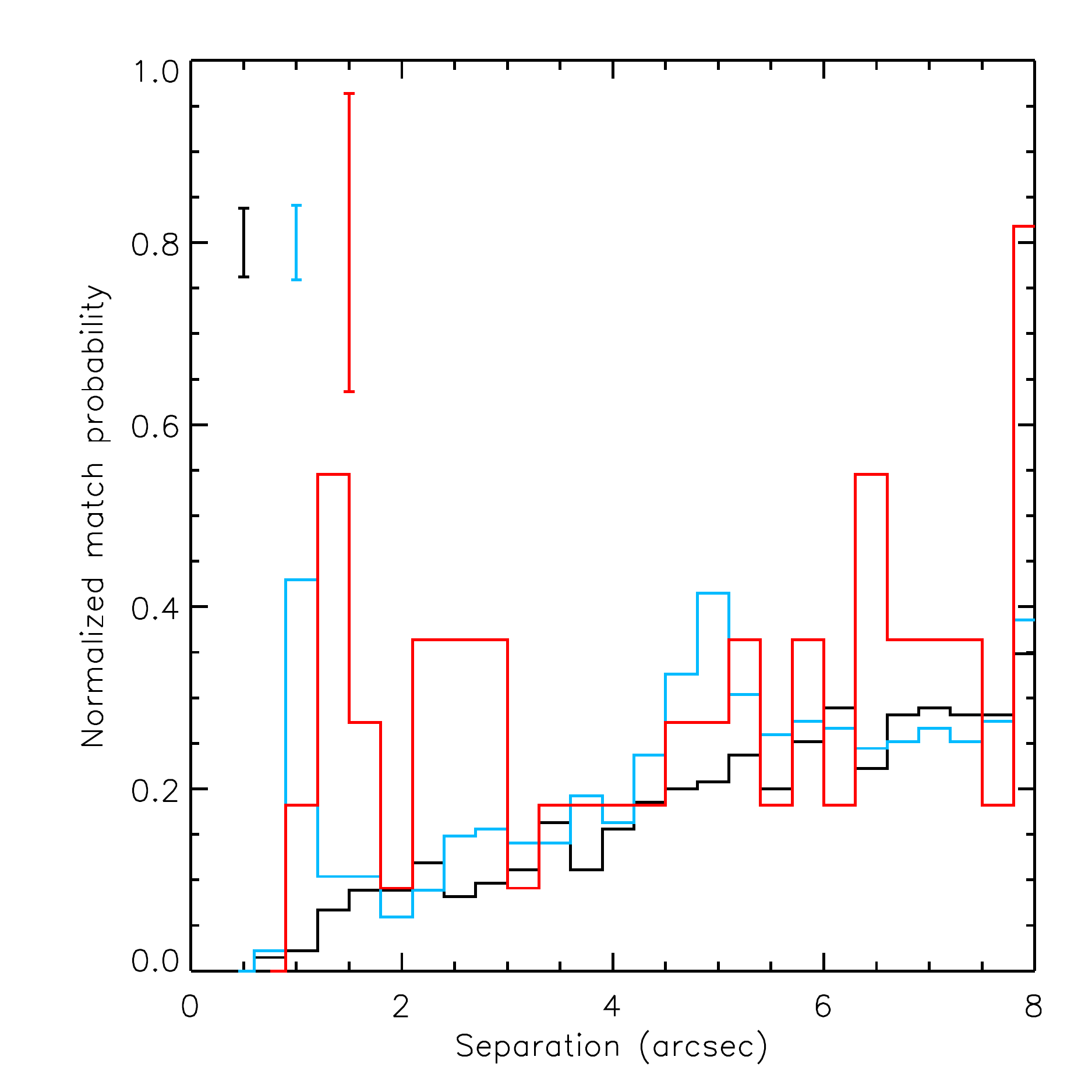}{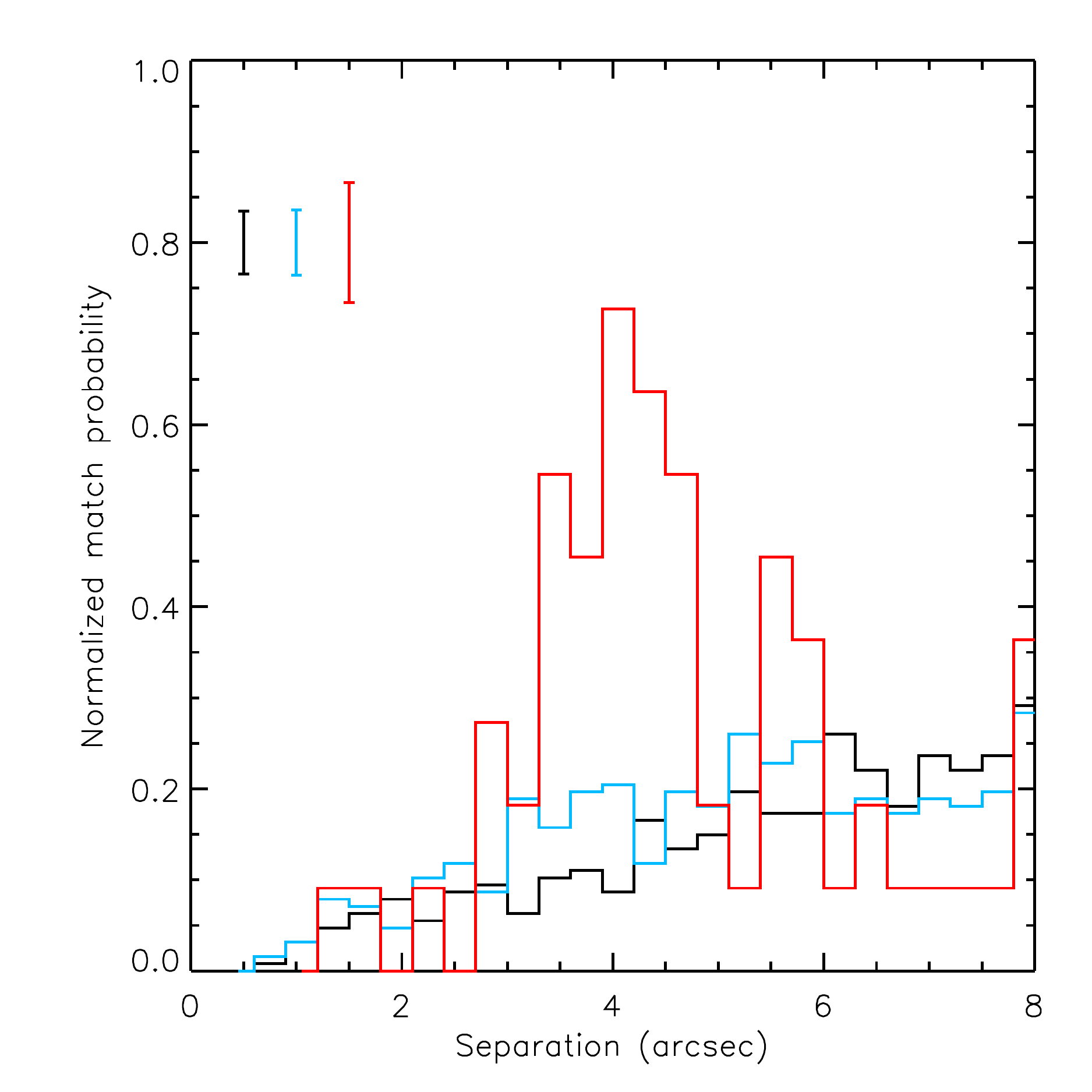}
\caption{Histogram of separations between red quasars and nearby sources detected in our SExtractor catalogs (red line).  
All three histograms are normalized to the number of input sources to represent a match probability.  We compare these with the separation histogram for morphologically stellar sources in each field (blue line) and chance coincidence matches to a false catalog created by shifting the morphologically stellar sources to the north by 15\arcsec.  We plot at the top left the mean size of an error bar  that would be centered on a bin for each population.
The distribution for sources in the F160W filter is shown on the {\em left}.  
Because of the smaller number of frames in each filter we combine the F105W sources with the F125W sources in the {\em right} hand panel.   In both the red and the blue images, red quasars show a significant excess of companions within 4\farcs5 of the quasar.  }\label{fig:sep_hist}
\hspace{0.5in}
\end{figure*}

We use the SExtractor catalogs to determine whether the excess of sources seen near our quasars is significant compared to the overall distribution of sources in each field.  Figure \ref{fig:sep_hist} shows histograms of the cumulative number of matches as a function of source separation in 0\farcs3 bins for the red quasars (red line) in the red and blue filters (excluding self-matches within 0\farcs1).  We compare this with the histogram of separations for the morphologically stellar sources (blue line) that we determined by differential photometry as described above.  We exclude our quasars from the stellar source histogram.  In addition we measure the distribution of random coincidences (black line) shifting the positions of the stellar sources by 15\arcsec\ to the north and matching to the source catalog.  The mean size of the error for each set of matches is shown at the top right.  

We see that morphologically stellar sources show some excess matches compared to the random background, but that the red quasars show significant excess in both bands.  A two-sided KS test comparing the distribution of sources near the quasars to the overall source distribution (black line) as well as to the morphologically stellar sources in the F160W filter yields p-values of $2\times10^{-9}$ and $3\times10^{-10}$ respectively, allowing us to reject the null hypothesis with greater than $5\sigma$ confidence.  Furthermore, KS test comparing the overall source distribution and the morphologically stellar distribution yields a p-value of 0.4, implying that for those distributions the null hypothesis cannot be rejected.  Similar results are obtained for the blue filter data ($1\times10^{-7}$, $3\times10^{-9}$ and $0.6$, respectively).  The excess of sources within 4\farcs5 is most pronounced.  At $z\sim2$ 4\farcs5 translates to a projected distance of $\sim 38$ kpc, and we adopt this distance as the upper limit for considering a companion system in our modeling. 

\section{Image Decomposition}\label{sec:galfit}

\subsection{Construction of the Point Spread Function}

In order to model the quasars, a point spread function (PSF) is needed to provide a standard of how real point sources (stars or quasar nuclei) are represented in a drizzled image from a particular telescope. Previous studies of quasar host galaxies have devoted up to several orbits of {\em HST} time to obtain a deep, high signal-to-noise ratio image of an isolated star to represent the PSF \citep{Floyd04,Urrutia08}.   
In more recent, similar work \citep[e.g.,][]{Simmons12,Schawinski12,Schawinski11}, which examined the host galaxies of moderate luminosity X-ray-selected AGN in WFC3/IR data, the PSF was constructed from images of isolated stars across the field.  These studies and our own examination, found that the PSF does not vary strongly across the field\footnote{In addition, all our objects are observed within 10 pixels of each other in the same location on the chip. }.  Therefore, we created one PSF for the each filter using stars in drizzled images of our own data as well as from archived observations in the same filter. We selected from the MAST archive all observations taken since 2012 January 1 with the WFC3 imager, in the IR aperture and the filters F105W, F125W, and F160W that used the same 4-point dither pattern.  We chose only science observations and did not consider calibration data.  We visually inspected the results and selected for retrieval fields that showed a few isolated and bright, yet unsaturated, stars.  We then processed the images in the same manner as our science data using the same {\tt astrodrizzle} task, approximately doubling the size of our program data to create a high signal-to-noise PSF.   

We made $201\times 201$ pixel cutouts around eligible stars and produced a single mosaic image of all the stars.  We masked out any extraneous light sources and replaced the masked pixels with the approximated level of background noise so as to not interfere with the PSF fitting.

We implemented the steps for PSF creation prescribed by the DAOPHOT package in the Image Reduction and Analysis Facility (IRAF) suite.
We produced a list of the peak positions of the chosen stars in the mosaic using the {\tt daofind} routine.
To properly weight the stars in the PSF, we used the {\tt phot} function to estimate their magnitudes using an aperture of 3.0 pixels and a sky annulus  with an inner radius of 10.0 pixels and outer radius of 20.0 pixels. The {\tt pstselect} algorithm then selected the brightest stars that were sufficiently separated from other bright stars, using a {\tt psfrad} of 100 pixels and a {\tt fitrad} of 4 pixels.  The output from {\tt pstselect} was then fed into the final PSF-making routine, {\tt psf}, which verifies the chosen stars and fits a 0th order Gaussian to produce a final sampled PSF look-up table. 
Finally, {\tt seepsf} task converts the lookup table to a FITS image of the PSF whose image size is $201\times 201$ pixels.  

\begin{figure*}
\plotone{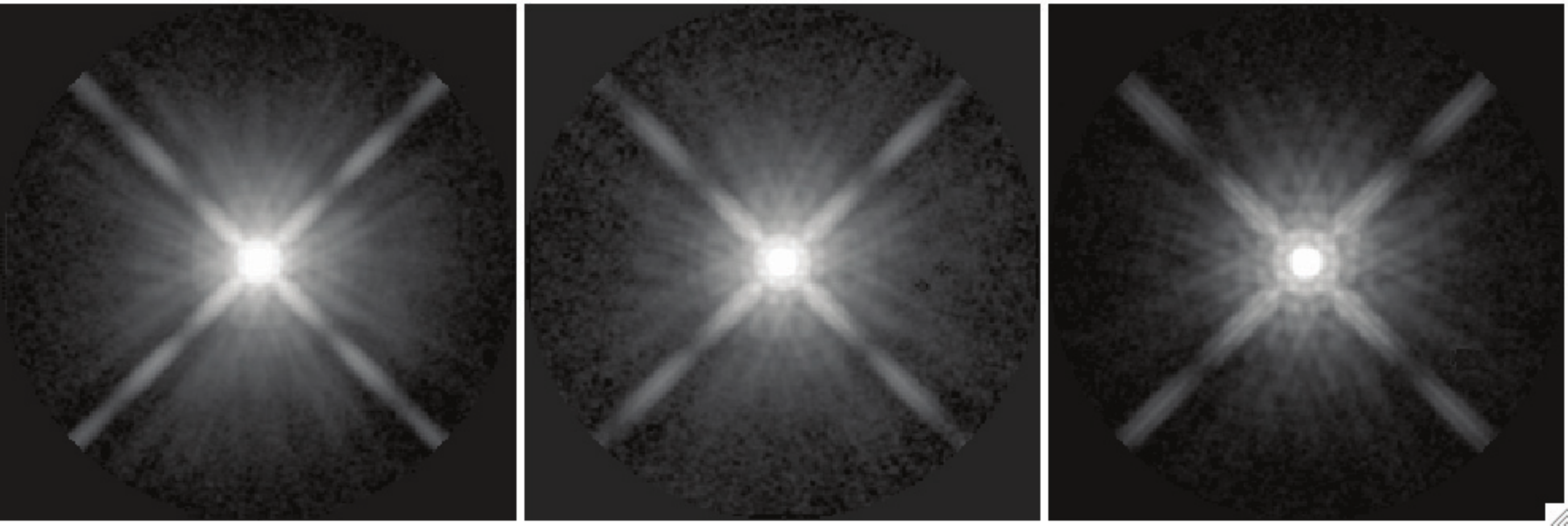}
\caption{Composite PSFs for the three filters used in our HST program. {\em Left --} F105W. {\em Middle --} F125W. {\em Right --} F160W.  The images are shown with a logarithmic scaling with intensity ranges set to include 99.5\% of the pixels and the image size is $201\times201$ pixels, corresponding to $12\farcs6\times12\farcs6$.}\label{fig:psf}
\hspace{0.5in}
\end{figure*}

Figure \ref{fig:psf} shows the final PSF images for the F105W, F125W, and F160W filters, respectively.  Table \ref{tab:psf} lists the relevant parameters for each PSF, including the number of archival fields used to supplement our proprietary data in column (2), the number of stars used to create the PSF in column (3), the PSF's full-width at half-maximum (FWHM) in column (4). 

\subsection{Multi-component fitting with GALFIT}\label{sec:fitresults}

Our objective is to study the morphologies of red quasar host galaxies and to determine whether mergers play an important role in their triggering.  To do this in a quantitative manner we carried out host/point-source decomposition using GALFIT \citep{Peng02,Peng10}, modeling our sources with a PSF plus as many host-galaxy components as necessary to minimize the reduced $\chi^2$ while careful not to overfit the data.  

We model the host galaxy components with a S\'{e}rsic radial profile \citep{Sersic68} given by the equation:
\begin{equation}
\Sigma(r) = \Sigma_e \exp\Bigg[ -\kappa \bigg( \Big(\frac{r}{r_e}\Big)^{1/n} -1 \bigg) \Bigg], 
\end{equation}
where $r_e$ is the effective radius within which half the total flux is contained and the surface brightness at the effective radius is $\Sigma_e$.  The parameter $n$ is referred to as the `S\'{e}rsic index' and determines the concentration of the light profile.  A profile with $n=4$ represents the light distribution of a classical bulge, while $n=1$ is an exponential disk which fits a classical disk.  When $n=0.5$ the function is a two-dimensional Gaussian profile.  The parameter $\kappa$ is tied to $n$, so that the S\'{e}rsic  index is the sole parameter that independently determines the radial light profile.  

GALFIT fits a S\'{e}rsic profile to an image by adjusting the following parameters and minimizing $\chi^2$ using the Levenberg-Marquardt algorithm: the $x$ and $y$ position of the profile's center; the source's integrated magnitude; $r_e$; $n$; the axis ratio $(b/a)$ and position angle.  

Previous quasar host studies have used a similar approach with independent software \citep{McLure00,Floyd04,Urrutia08} that fits separate PSF and \sersic\ components in a two step process, where the PSF subtraction is performed initially by scaling the PSF to the peak flux in the quasar, followed by a host-galaxy fit. 
However, subtracting the point source first can bias the host galaxy fit, while GALFIT performs the point source plus additional component fitting simultaneously, reducing this bias. 

To begin fitting morphological components to our images, we extracted a $201\times201$ pixel box centered on quasar's peak, as shown in Figures \ref{fig:hst_images} and \ref{fig:hst_images2}. Since GALFIT aims to fit all the flux in an image, it is important to mask out any additional sources of light or bad pixels in the image.  We show the masked regions in the residual (4th) column of Figures \ref{fig:hst_images} and \ref{fig:hst_images2}. 

Because reddening dims the quasar at shorter wavelengths, the host-to-point-source flux ratio is larger in the F105W and F125W filters compared with the F160W filter.  Therefore, to better decompose the two systems without the point source overwhelming the host flux, we performed the first fit for each source in the shorter wavelength filter. We then fit the F160W images independently, but informed by the results of the shorter wavelength fits\footnote{We experimented with fixing the parameters of the F105W/F125W components in the F160W images, but found that these often did not converge or yielded poorer results, with larger $\chi^2$ values.}.  

Our ultimate goal is to remove the point source and recover flux from the underlying host galaxy, whose central position would be within 0\farcs6 as the PSF, as well as any nearby companions.  We define `nearby companion' as any component with a separation between 0\farcs6 and 4\farcs5 arcsec, which translates into a projected distance of $\sim 5$ to $\sim 40$ kpc at the redshift of our quasars.  However, because an underlying host galaxy can be elusive,
we began each fit with a single PSF plus S\'{e}rsic index centered on any obvious companion galaxy component, plus a flat sky.  If no obvious component was visible, we initially fit just a PSF plus sky.  We then added an additional host galaxy S\'{e}rsic component at the location of the quasar and compared the reduced $\chi^2$ value with the added component to decide whether an additional component was warranted.  We adopted an added component if the reduced $\chi^2$ was significantly improved ($>5\sigma$ in an F-test). 

In many cases, adding a \sersic\ component at the same location of the PSF in order to model the underlying host galaxy resulted in a significantly improved fit, but with GALFIT assigning unphysical parameters that imply a need to fit flux from an unresolved region at the position of the PSF (i.e., $r_e \lesssim 3$ pixels and $n \lesssim 0.1$).  While unphysical, this added component improves the $\chi^2$ likely because it captures the residual noise from the single model PSF fit.  However, since we are interested in decomposing the PSF from any discernible underlying host, in cases where an unphysical \sersic\ profile arose, we tried a fit with two PSF components whose position and magnitude GALFIT could adjust to account for residual flux from a single PSF fit.  In all but two cases where a second PSF was attempted, the fit was significantly better (as determined by an F-test which gave $\sigma > 10$) and allowed for a more physical underlying host galaxy to be fit by an additional \sersic\ component.  Although we use two PSF components to accommodate the flux from the quasar, we do not interpret this as evidence for a dual AGN, since the separations between the two components are smaller than the spatial resolution of the images and are more likely due to residuals. The difference in PSF magnitudes between the two components was generally smaller for the blue bands, ranging between 0.2 and 1.5 magnitudes, compared with a range of 0.5 to 3 magnitudes for the F160W images.

To better understand the nature of the fits needing two PSF components, we tested this technique on ten bright stars found in five of the quasar fields.  The brightness range of the stars spanned 6 magnitudes.  The brightest stars that we tested were two magnitudes brighter than the brightest quasar in our sample, and the faintest star that we tested was two magnitudes fainter than the faintest quasar (the quasars themselves span about 2 magnitudes in brightness).  We found, consistently, that single-PSF fits to the brighter stars resulted in large, symmetric residuals and high reduced $\chi^2$ values ($>20$ for the two brightest sources) which were improved significantly (by $\gtrsim30\%$) with the addition of a second PSF component, but never succeeded at capturing all the flux. The fainter stars generally yielded good fits (typical $\chi^2 < 3$) with a single PSF and were not improved significantly with the addition of a second component. 

We combine the fluxes from the two PSF components into a single PSF magnitude in the following way: 
\begin{equation}
m_{\rm Tot} = -2.5 \log_{10}( 10^{-0.4 m_{\rm PSF_2}} + 10^{-0.4 m_{\rm PSF_2}} ).\label{eqn:phot_2psf}
\end{equation}
In the two cases where an unphysical \sersic\ profile best fits the image, we ascribe the flux in the \sersic\ component to be part of the point source, combining the PSF magnitude ($m_{\rm PSF}$) and \sersic\ magnitude ($m_{\rm S}$) provided by GALFIT similarly:
\begin{equation}
m_{\rm Tot} = -2.5 \log_{10}( 10^{-0.4 m_{\rm PSF}} + 10^{-0.4 m_{\rm S}} ).\label{eqn:phot_psf}
\end{equation}
We report these combined magnitudes as the PSF magnitude in Table \ref{tab:galfit_params} and indicate the origin of the magnitude with a footnote in the table.  We do not show the second components in the galaxy model represented in the third column of Figures \ref{fig:hst_images} and \ref{fig:hst_images2}.  

In many cases, additional S\'{e}rsic components are needed to better fit the companion systems, and those components may not have the same parameters, or even location in the two bands.  We interpret these differences as potential regions of star formation which are bright below 4000\AA\ (in the rest frame) showing up in the blue band, but not in the red band, or luminous regions of dusty star formation that would appear red rather than blue. While we can extract physical information about the companion systems from their Galfit parameters, in many of the sources the S\'{e}rsic index that provides the best fit may not be physically meaningful (i.e., $n < 1$ or $n>4$).  Likewise the effective radius can range as high as 200\arcsec.  
 
From Figures \ref{fig:hst_images} and \ref{fig:hst_images2} we see that all but two (F2M0943 and F2M2222) sources show nearby companions to the central quasar with disturbed morphologies.  Another source (F2M1359, see Section \ref{sec:f2m1359}) is fit by a relatively smooth and centrally located galaxy and may also not be hosted by a merger or even intrinsically reddened.  
 
 \subsection{Uncertainties in the multi-component fits}\label{sec:uncert}
 
 Although the \sersic\ profile may provide some physical insight into the light distribution of a galaxy, there can be multiple components and tidal tails that are not well approximated by the shape of the profile leading to large errors in the fitting parameters.  Here we discuss the uncertainties of our results, keeping in mind that the uncertainties reported by GALFIT represent only the errors from the covariance matrix and do not generally account for other sources of error; reporting only these sources of uncertainty may underestimate the true errors. However,  there have been multiple studies characterizing the additional uncertainties in multi-component galaxy fits in the presence of a bright AGN. The first and largest of these studies was performed by \citet{Simmons08} who simulated more than 50,000 images of AGN and host galaxies with a range of galaxy and nuclear properties. 
 
\citet{Simmons08} conclude that recovery of AGN and host galaxy characteristics is generally very reliable. AGN and galaxy properties are accurately recovered even in automated batch-mode fits, including cases where the host galaxy is outshined by the nuclear point source. In individually fitted images where the specifics of each source and image may be properly accounted for (e.g., companion galaxies and stars, noise variances across the image), the uncertainties are reduced from the batch-mode case. 

The uncertainties in component \sersic\ indices are also reduced when the centroids of the bright point source are separated from those of the galaxy components by at least the FWHM of the PSF, as is the case for all the companion systems in this work. Because the error tables in \citet{Simmons08} assume positionally coincident components and batch-mode fitting, their uncertainty values are high compared to what we expect for the true uncertainties in this work. Nevertheless, the predicted conservative uncertainties are useful as guidelines here. The error tables predict typical uncertainties in AGN brightness of 0.25 mag, typical errors in host brightness of 0.5 mag, and typical errors in host \sersic\ index of $\delta_n \simeq 0.7$. In most cases this does not affect the assessment of a galaxy component as likely to be disk-like or bulge-like. Host galaxy sizes ($r_e$) are more uncertain and the uncertainty depends somewhat on the fitted morphology.  We list in Table \ref{tab:galfit_params} uncertainties for the underlying host parameters as a combination of reported errors by GALFIT and estimated additional uncertainties due to multi-component AGN and host galaxy fits from the simulations of \citet{Simmons08} added in quadrature.  The companion systems are far enough away from the PSF that we do not expect the AGN to affect their fitting errors and we list for them the uncertainties that are reported by GALFIT.

The error tables in \citet{Simmons08} assume no detailed follow-up from batch-mode fitting is performed and estimate maximum errors due to positionally coincident AGN and host galaxy centroids, so their use in this study produces errors estimates that are conservative with respect to the true uncertainties. 
While an analogous study to \citet{Simmons08} basedon WFC3/IR data at our depth and AGN to host ratios would be the ideal method for estimating our uncertainties, such an effort is significant and beyond the scope of this paper.  
We use the \sersic\ components primarily as a means of capturing all the flux in the images so that we can study the colors and luminosities of the merging components to better understand the co-evolution of merging galaxies hosting luminous red quasars (see Section \ref{sec:galaxies}).  

\section{Results}\label{sec:results}

\begin{figure*}
\plottwo{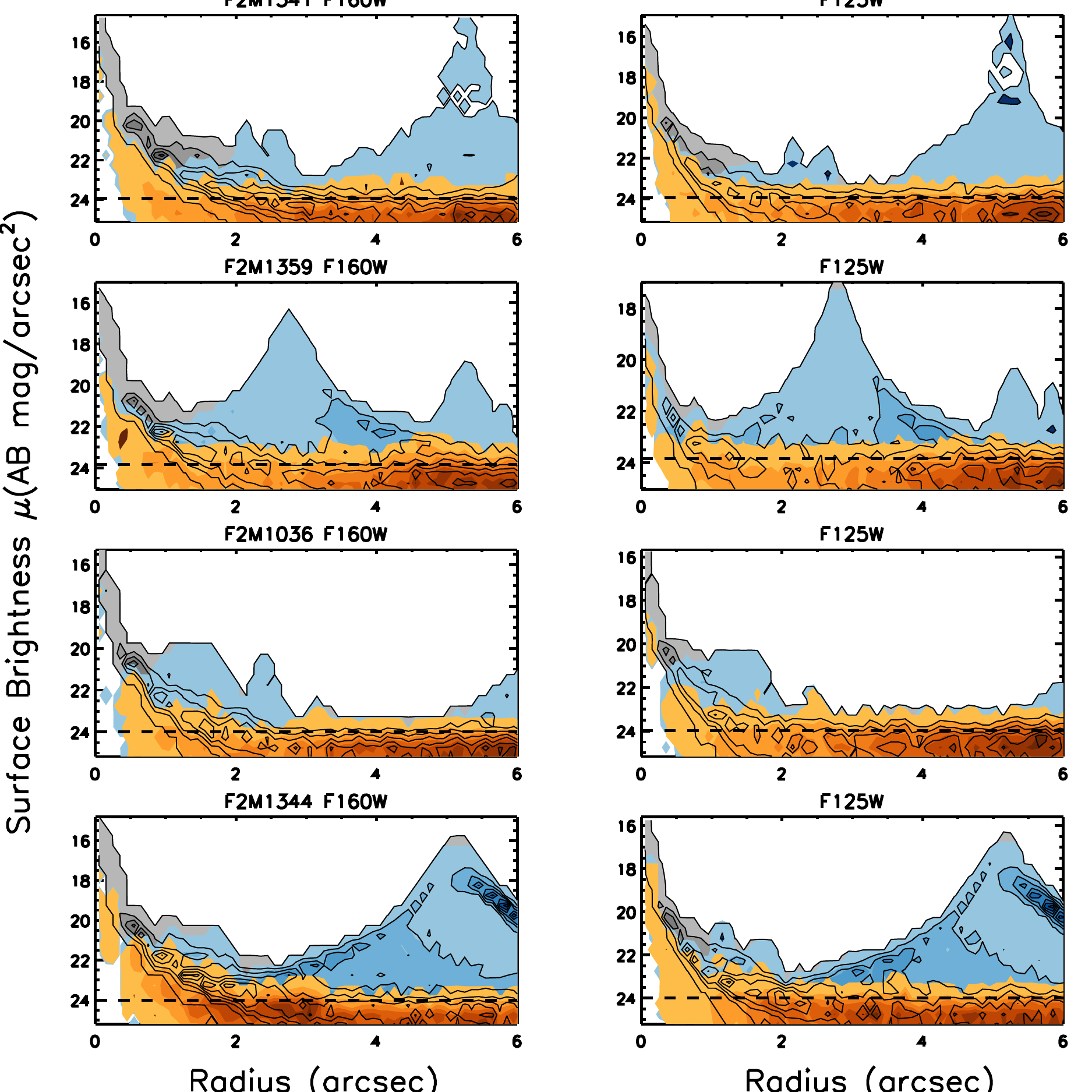}{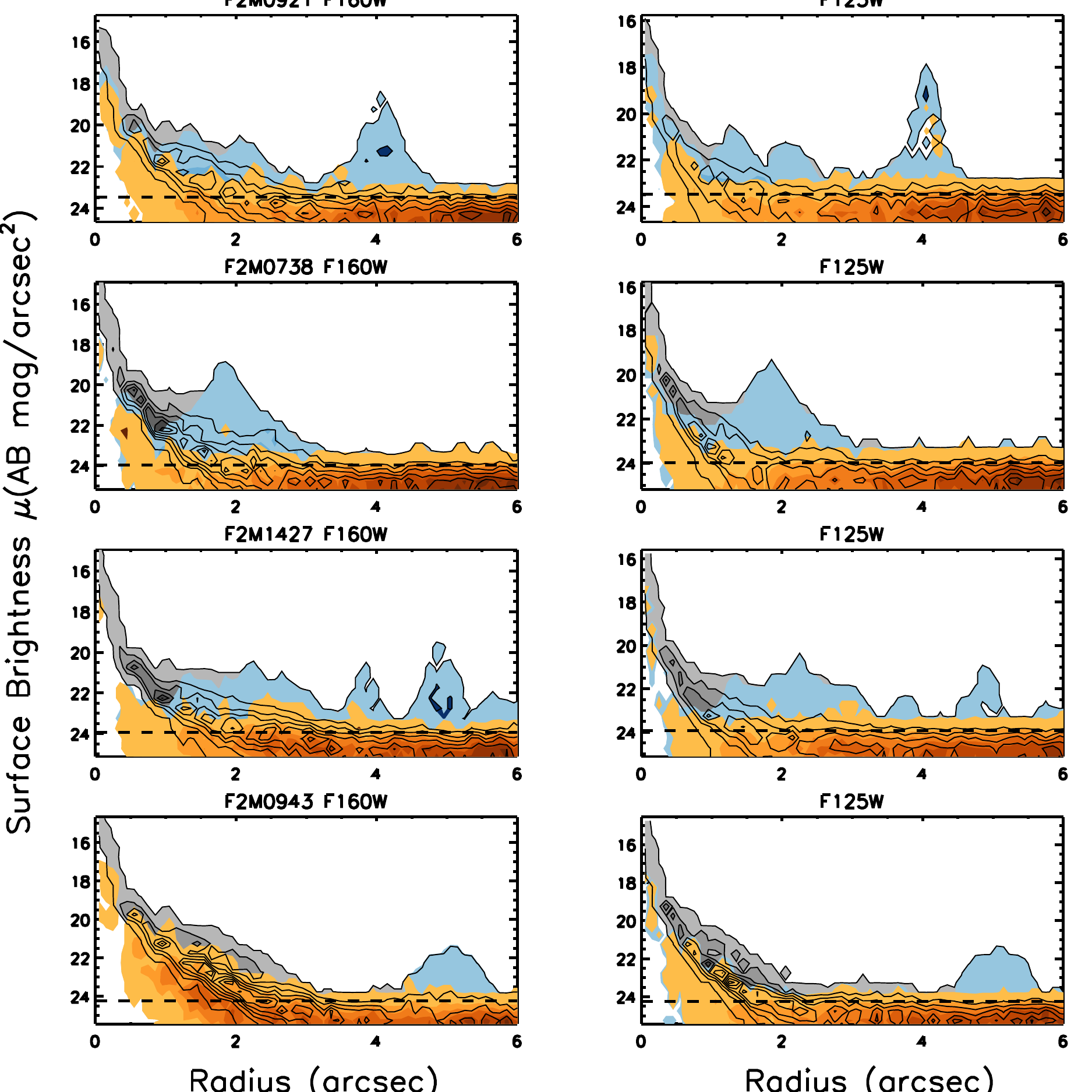}
\caption{Shaded contours show the surface brightness in each pixel as a function of distance from the point source (radius) in arc seconds for the eight quasars in Figure \ref{fig:hst_images}.  The panel on the left for each object plots F160W data while F105W is shown on the right.  Grey shaded regions and black contours show the flux from the unaltered quasar image, while the blue regions show the flux distribution from the PSF-subtracted image, and the orange regions represent the data from the masked residual images.  The 3$\sigma$ threshold is shown with a dashed horizontal line.  In each panel, with the exception of F2M0943, significant structure  (i.e., companion galaxies or tidal features) is seen immediately beyond the PSF.  }\label{fig:profile}
\hspace{0.5in}
\end{figure*}

\begin{figure}
\plotone{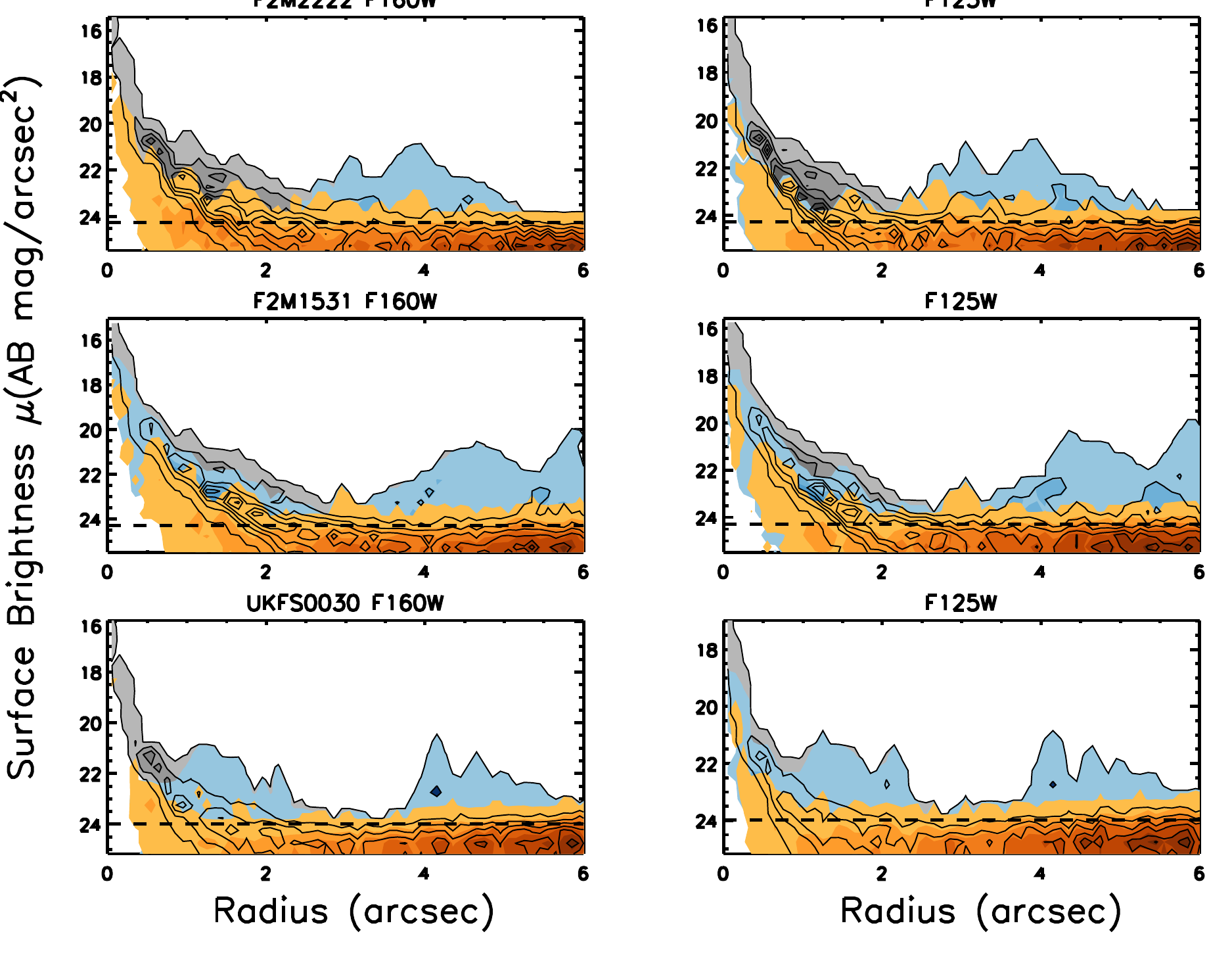}
\caption{Surface brightness as a function of radius in arc seconds the three quasars shown in Figure \ref{fig:hst_images2}.  The panel on the left for each object plots F160W data while F125W is shown on the right. Contours are colored the same as in Figure \ref{fig:profile}. }\label{fig:profile2}
\hspace{0.5in}
\end{figure}

\begin{figure}
\plotone{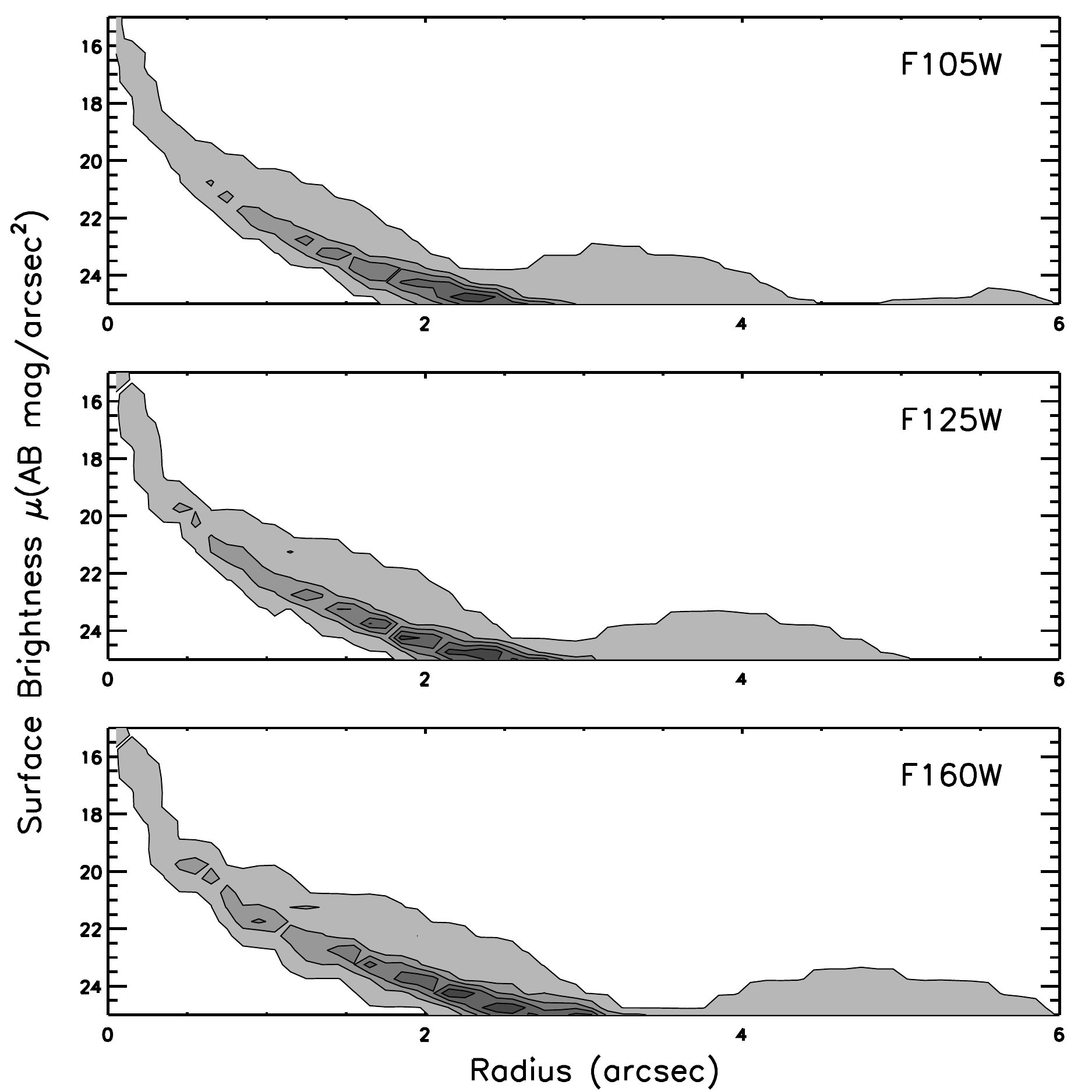}
\caption{Surface brightness as a function of radius in arc seconds for the three PSF images (Figure \ref{fig:psf}).}\label{fig:psf_profile}
\hspace{0.5in}
\end{figure}

With the fully reduced, PSF-subtracted and residual images in hand, we can study their surface brightness distribution as a function of distance from the central point source.  Because the red quasars' images are highly asymmetric (see \S \ref{sec:nonparam}) the single radial profile that is typically plotted for galaxy light distributions is insufficient to describe the profile of these red quasar hosts.  We plot in Figures \ref{fig:profile} and \ref{fig:profile2} the surface brightness distribution, $\mu$, in mag arcsec$^{-2}$ for all the pixels in all three images as a function of radial distance from the quasar's peak emission in each filter. The gray shaded regions and black contours represent the flux in the original image, while the blue contours are the fluxes from pixels in the PSF-subtracted image\footnote{By `PSF-subtracted' we mean the removal of all centrally concentrated light, including from a second PSF component or a concentrated \sersic\ component (e.g., F2M1341)}.  We plot the residual image's flux in orange, and use their values to determine the statistical limit of our observations.  We compute the standard deviation of the flux for all pixels in the masked residual image (as shown in the fourth column of Figure \ref{fig:hst_images}) using the IDL procedure {\tt mmm} which is part of the astronomy routine library.  The horizontal dashed line represents this 3$\sigma$ limit and is quoted in Table \ref{tab:qso_sample}.  For comparison, we plot in Figure \ref{fig:psf_profile} the same for the PSF profiles.

In all the panels, except for F2M0943, we see significant structure beyond the point source.  F2M0943 is the one object that does not favor a merger. Its profile in both filters are similar to the PSF profiles, as shown in Figure \ref{fig:psf_profile}. 

The dynamic range of our images, which we define as the ratio of the peak flux in the point source to a 5-$\sigma$ detection threshold ($D=N_{\rm PSF}/N_{5\sigma}$), is between 960 and 4000 in the red (F160W) band. In the blue bands (F105W and F125W) the dynamic range is between 360 and 4500.  The large dynamic range of our images poses the biggest challenge for our ability to study the low surface brightness host galaxy light after point source subtraction.  This is because the poisson noise in the point source goes as $\sqrt{N_{\rm PSF}}$, which implies that the residuals after PSF fitting and subtracting is on the order of $\sqrt{N_{\rm PSF}}$.  When $D$ is large, then $\sqrt{N_{\rm PSF}} >> N_{5\sigma}$, which ranges between 160 and 370 for the red filter and 60 to 340 in the blue filter.  Residuals tens to hundreds of times brighter than the faint host galaxy features means that even with perfect PSF modeling the Poisson errors will dominate over the underlying galaxy.  In our sample, the source with the largest dynamic range in both filters is F2M0943.  

Most of the systems studied in this sample show evidence for mergers hosting the red quasars, although their details are heterogeneous and complex. We note that in the redshift range of our sample ($z=1.7-2.3$) many of the tidal features seen in the $z\sim 0.7$ sample from \citet{Urrutia08} would disappear.  This was shown by \citet{Schawinski12} who simulated the appearance of four F2M quasar from \citet{Urrutia08} in the WFC3 $H$ band when redshifted to $z=2$.  Nevertheless, some merger signatures are still evident in the images.  We approach the interpretation of our images below with this result in mind, cautious not to over-interpret the details of the fitted parameters.  

\subsection{Red Quasar Properties}\label{sec:qso}

\citet{Urrutia08} found that once the point source was separated from the host galaxy, the quasars themselves (i.e., the fitted PSFs in that sample) had {\em redder} colors than the low-resolution photometry reported for the systems as a whole.  In fact, the shift in color for these objects was larger for sources with redder total $E(B-V)$ values (as determined from spectral fitting).
They attribute this effect to an excess of blue light in the F475W filter, which is bluer than rest-frame 4000\AA\ break. \citet{Urrutia08} interpret this as an excess of young stars adding blue light from the host to the total integrated colors.  

Do we see a similar situation at $z\sim 2$, where the quasars are more luminous by $\sim 2-3$ magnitudes, on average, than the $z\sim 0.7$ sample?  
We use the SExtractor catalogs described in Section \ref{sec:obs} as well as the magnitudes from the Galfit modeling to examine the effect of separating the host galaxy light on the PSF colors. 
In six cases we detect the underlying quasar host galaxy and detect offset components in the other three cases.
 Our experimental design was intended to measure the same rest-frame emission as in the \citet{Urrutia08} work, with filters chosen to straddle the 4000\AA\ break.  The F160W filter corresponds to $\sim 4850 - 5890$\AA\ in the rest frame, depending on the redshift, while the F105W filter corresponds to $\sim 3250 - 3870$\AA\ and the F125W filter corresponds to $\sim 3790 - 3840$\AA. We therefore discuss the colors of the quasars in terms of rest-frame $U-V$ corresponding to the F160W filter and either F105W or F125W.  

\begin{figure}
\plotone{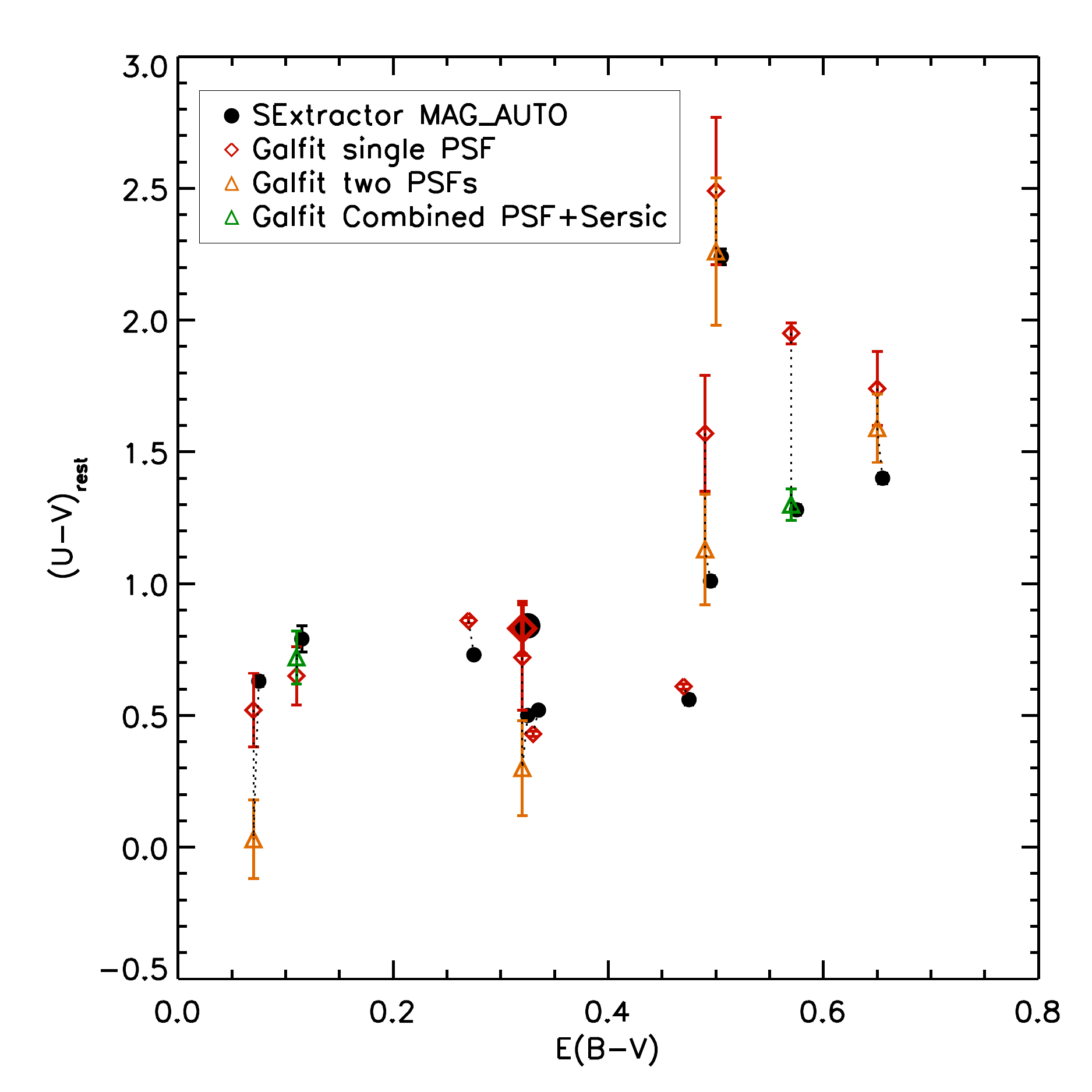}
\caption{We plot rest-frame $U-V$ color of red quasars versus $E(B-V)$ to investigate whether the removal of host galaxy light results in a redder quasar component, as was found in the $z\sim 0.7$ sample of \citet{Urrutia08}.  The filled circles are from the total magnitudes estimated by SExtractor's {\tt MAG\_AUTO}  (offset to the right by 0.005 magnitudes, for clarity), which we compare with the PSF magnitudes determined by Galfit (red diamonds). In sources that are better-fit by a second PSF component 
we plot with orange triangles the combined flux as computed by Equation \ref{eqn:phot_2psf}. While sources needing an added \sersic\ component to absorb residual PSF flux are shown with green triangles computed by Equation \ref{eqn:phot_psf}.  When taken alone, the PSF component is typically redder than the total source magnitude, as well as the combined magnitude from an added component consistent with the lower redshift sample.  This may indicate the presence of blue light added by a nuclear starburst,  but could also be an indication of the inadequacies of the single PSF fits.  Two quasars (F2M1531 and UKFS0030) have the same reddening, with $E(B-V)=0.32$.  To distinguish between them, we plot UKFS0030 with enlarged symbols. }
\label{fig:psfcolor}
\end{figure}

We plot in Figure \ref{fig:psfcolor} the rest-frame $U-V$ color of our quasars as a function of their reddening, $E(B-V)$ from  \citet{Glikman12}, comparing the total color of the source, as measured by the the {\tt MAG\_AUTO} parameter in SExtractor, to the magnitudes returned by the Galfit modeling. SExtractor's {\tt MAG\_AUTO} parameter is the equivalent of a ``total magnitude'' encompassing $>90\%$ of the total contiguous flux centered on the peak of the light profile and therefore includes the quasar plus the host.  We plot the $U-V$ color from the SExtractor photometry with black circles.  
Red diamonds show the PSF components' colors, while green triangles show the combined PSF component plus \sersic\ component for sources that are better-fit by such an added component. 
The dotted lines connect the values for a given source to guide the eye. 

In general, quasars with higher $E(B-V)$ values also have larger $U-V$ colors.
Although in most cases (6/11) the PSF-only component does have redder colors than the total {\tt MAG\_AUTO} magnitude or the PSF plus \sersic\ component, we do not see the clear trends that \citet{Urrutia08} noted in the $z\sim 0.7$ sample.  
The primary PSF component is redder than the combined PSFs or PSF plus \sersic\ model and may contain some blue host flux as was found in the $z\sim0.7$ sample.  
However, because the dynamic range is higher in the $z\sim 2$ sample, compared with the $z\sim 0.7$ sample, and since the WFC3/IR spatial resolution is lower than the ACS resolution, separating the point source from the host galaxy for these systems is more challenging; we cannot say with certainty whether or how much of the additional blue flux in the added components is due to the quasar or young stars in the host.  

Consistent with the notion that merger-induced star formation adds blue light to the host, we note that the two sources with the smallest color difference between the different photometric measurements, F2M0943 and F2M2222, do not appear to have nearby companions or be actively merging.  Table \ref{tab:psf_mags} lists the magnitudes and colors of the quasars using the three metrics shown in Figure \ref{fig:psfcolor} (see \S \ref{sec:indiv_qsos} for details on individual sources).

\subsection{Host Galaxy Properties}\label{sec:galaxies}

Having separated the quasar emission from the underlying galaxy light, we can explore some of the host galaxy properties and compare them to what is known about normal and star-forming galaxies at similar redshifts. We add up all the flux that Galfit assigns to the non-quasar \sersic\ components by summing the magnitudes in flux space, in a similar manner used for the point sources.  We compute the total magnitude of the host plus companion galaxy in each filter following,
\begin{equation}
m_{\rm Tot} = -2.5 \log_{10}( \sum_{i}^{N_{\rm Ser}} 10^{-0.4 m_{\rm i}} ),\label{eqn:phot_serc}
\end{equation}
where $N_{\rm Ser}$ represents the number of \sersic\ components in our best fit.  

\begin{figure}
\plotone{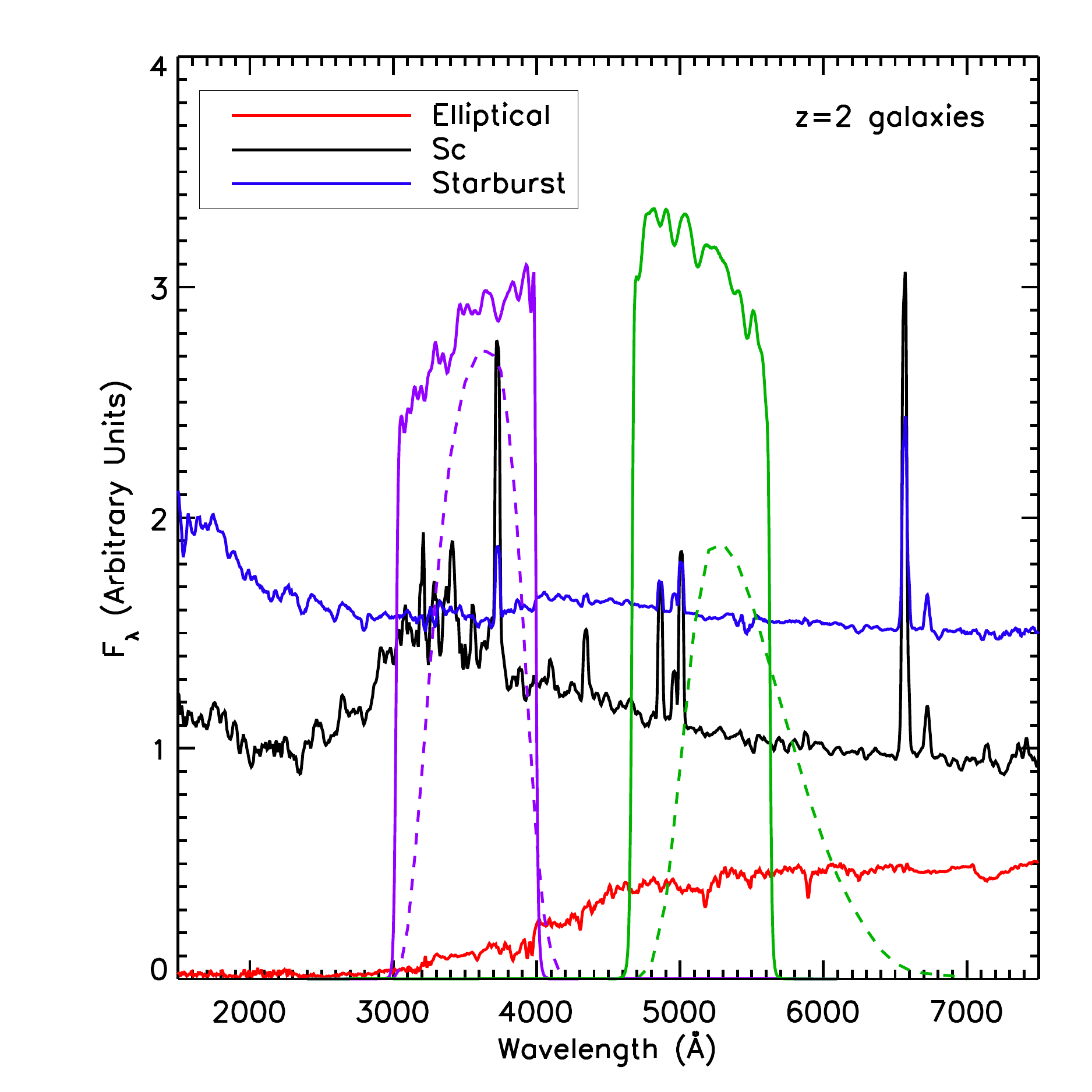}
\caption{ 
Comparison of three galaxy spectra from \citet{Kinney96}, namely, elliptical (red), Sc spiral (black), and starburst (blue) to the WFC3 F105W (solid purple line) and F160W (solid green line) filters at $z=2$, which correspond to rest-frame $U$ (dashed purple line) and $V$ (dashed green line) filters.}
\label{fig:uv}
\end{figure}

Our chosen WFC3/IR filters closely correspond to rest-frame $U$ and $V$ bands at $z=2$.  This is illustrated in Figure \ref{fig:uv} where we plot three rest-frame galaxy spectral templates from \citet{Kinney96}: Elliptical, Sc and a moderately reddened starburst template with $0.39<E(B-V)<0.50$, chosen to span a wide range in star-formation rates.  We plot the Johnson $U$ and $V$ filter curves in dashed purple and green lines, respectively, showing the sampling of the SEDs at these wavelengths. We then shift the F105W and F160W bandpasses to the rest-frame at $z=2$ and plot them with solid purple and green lines, respectively.  The same can be done for the three highest redshift sources, where the F125W filter corresponds to the $U$ band.  It is evident from this figure that the observed IR and rest-frame UV/optical curves overlap significantly, allowing us to compare the red quasar host colors with the $U-V$ colors of galaxies across the Hubble sequence at comparable redshifts from the literature.  

\begin{figure*}
\plottwo{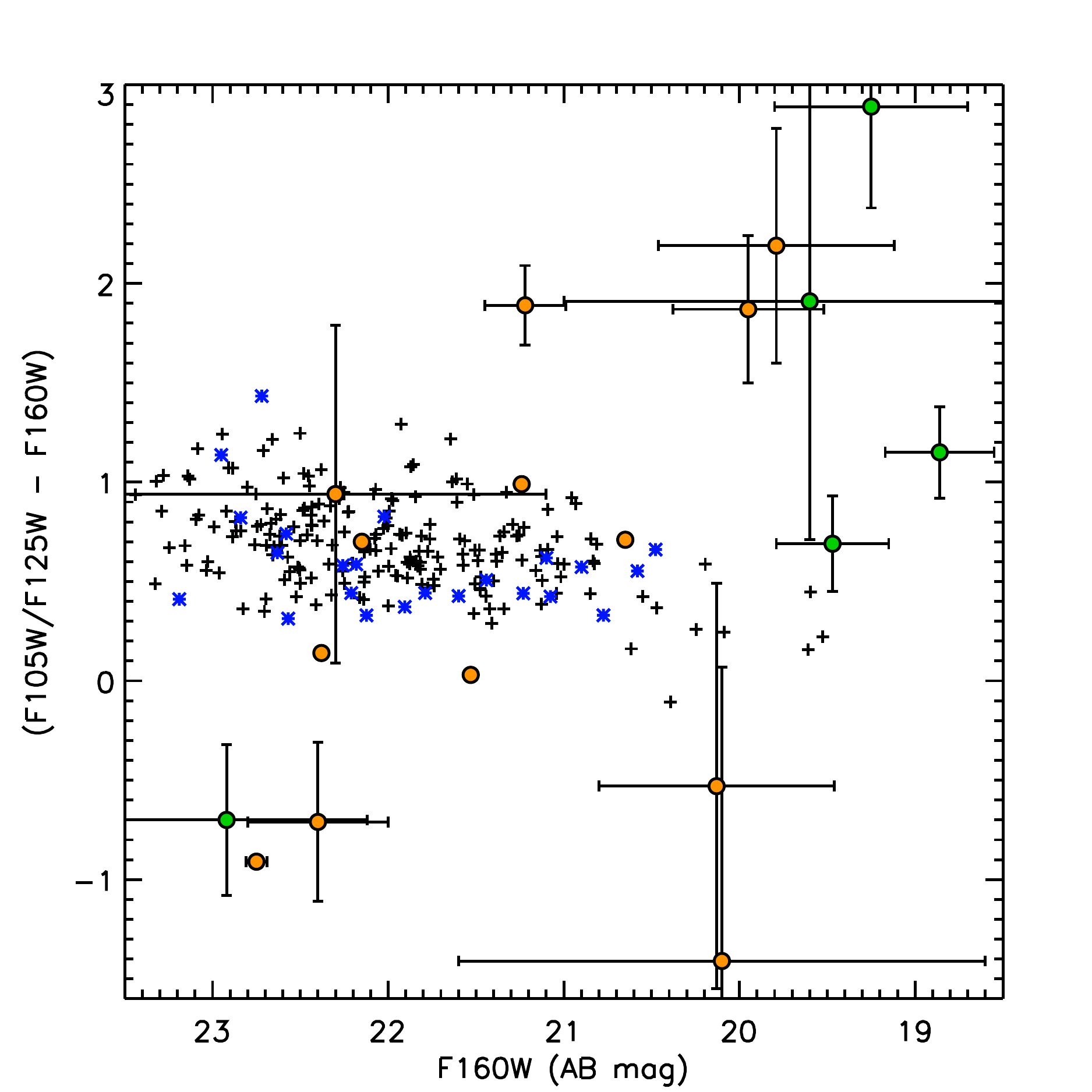}{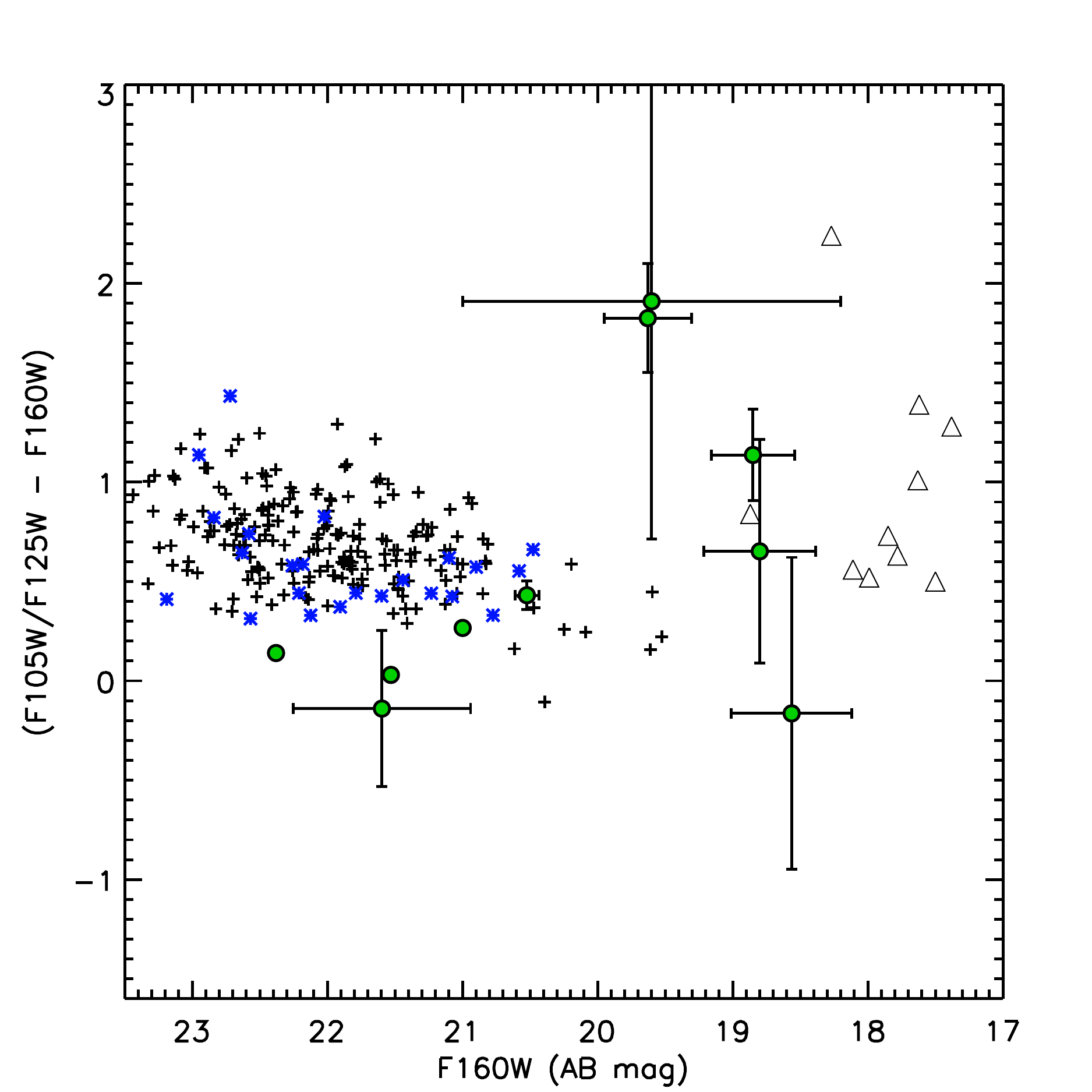}
\caption{ {\em Left --}  We plot the observed F105W$-$F160W (or F125W$-$F160W when applicable) as a proxy for rest-frame $U-V$ color versus the observed F160W magnitude of red quasar host galaxies (green circles) and companion systems (orange circles). For comparison we plot the same for galaxies between $1.7 < z <2.3$ from the CANDELS survey \citep[black crosses][]{Bell12}, while blue asterisks are AGN in the GOODS field from \citet{Simmons12}.    {\em Right --} F105W$-$F160W (or F125W$-$F160W) versus rest-frame F160W magnitude determined by summing all the components associated with the quasar (green circles). Black crosses and blue asterisks have the same meaning as in the left hand panel.
Open triangles are the total magnitudes of the sources ({\tt MAG\_AUTO}), plotted for comparison, to indicate the extent to which imperfect point source modeling may affect the galaxy colors.}
\label{fig:uv_mv}
\end{figure*}

In Figure \ref{fig:uv_mv} we plot the observed infrared colors of the red quasars as a proxy for rest-frame $U-V$ versus F160W magnitudes (corresponding to rest-frame absolute $V$-band magnitude) of the red quasar hosts (green circles) and companions (orange circles).  For a comparison sample, we turn to the Cosmic Assembly Near-infrared Extragalactic Legacy Survey \citep[CANDELS;][]{Grogin11,Koekemoer11} which consists of deep, multi-cycle observations of well-studied legacy fields with the {\em HST} WFC3/IR camera using the same three filters as in this work. \citet{Bell12} and \citet{Lee13} have studied the morphological and star forming properties of CANDELS galaxies out to $z\gtrsim2$ and provide a useful comparison set to our red quasar host galaxies.  We plot with plus signs CANDELS galaxies that had matches to sources in the publicly released catalogs of the UKIDSS Ultra-deep Survey \citep[UDS; from][]{Galametz13} and whose photometric redshifts are between $1.7 < z <2.3$ and stellar mass $M_\star > 3\times10^{10} M_\odot$, which \citet{Bell12} states as their completeness limit.  In addition, we plot with blue asterisks CANDELS photometry in the GOODS-South field of AGN in the same redshift range from \citet{Simmons12}.  We see that the AGN and normal galaxies lie in the same part of this parameter space.

The nearby companions to the red quasars (i.e., the orange circles) with the smallest photometric uncertainties lie in the locus of CANDELS galaxy colors. However,  four out of the five host systems (i.e., the green circles) in the left hand panel are among the most luminous galaxies.  The magnitudes of these components may be contaminated by imperfect point source modeling and therefore their luminosities may be overestimated. However, since their magnitudes are computed from the model fits, our estimates in most cases exclude residual host galaxy light in, e.g., clumps of star formation, which the smooth \sersic\ fit does not account for, which could lead to a lowering of their estimated luminosities.  

The  very high luminosities of the red quasar hosts is not unexpected given the relative volumes probed by the CANDELS and F2M/UKFS surveys.  CANDELS is a very small-volume survey covering a total area of $\sim 800$ arcmin$^2$, while FIRST, 2MASS and UKIDSS are all very large-volume surveys, enabling us to find the most luminous sources such as these red quasars. Therefore we expect our objects to lie in the high-mass/luminosity side of the diagram. Since stellar masses have been computed for the CANDELS sources, we examine their stellar masses in the magnitude range in which our systems lie and find that they are all comparatively high mass systems, with $M_\star \gtrsim 3\times10^{11} M_\odot$. 

In the right hand panel, we plot the total magnitude of all the light from host plus companions for a given quasar (computed using equation \ref{eqn:phot_serc}) to examine what the total magnitude of the system might be under the assumption that all the components are part of a merger that will eventually coalesce.  Here, the colors of five of our red quasars appear to continue along the locus of CANDELS galaxy colors, even when including light from the nearby companions. 

The colors of the F2M quasars span a very broad range, broader than the CANDELS galaxies, which is indicative of the complex nature of these putatively merging systems, where dust extinction (leading to red colors) competes with star formation (leading to blue colors). If the CANDELS galaxies are undergoing secular evolution, their colors ought to change more gradually with mass. 
We know there is significant dust in the red quasar systems, since their intrinsically very luminous quasars are dust-reddened. For some of the companions with very blue colors there may be unobscured star formation, making the hosts extremely blue. In other hosts, the star formation is behind dust, making them very red.

To check whether imperfect PSF modeling affects the colors of the host galaxies, we plot with triangles the uncorrected total magnitudes ({\tt MAG\_AUTO} from the SExtractor catalogs) of the red quasars in the left hand panel of Figure \ref{fig:uv_mv}.   
The total colors of our sources are bluer than the colors of the galaxies, implying that the blue colors of their host galaxies might be contaminated by some contribution from the point source.
 
\subsubsection{Non-parametric Galaxy Properties}\label{sec:nonparam}

Our parametric approach of fitting \sersic\ profiles was largely intended for PSF subtraction and capturing the residual flux for separate analysis.  We are cautious not to over-interpret the fitted parameters themselves since parametric fitting methods are often insufficient to describe merging and irregular galaxies and fail to account for all the structure that is seen.  Non-parametric fitting techniques have therefore been developed to asses the degree of irregularity in galaxy images \citep{Abraham03,Lotz04}. The Gini coefficient, $G$ -- originally an econometric tool devised to asses the wealth inequality of a population -- has been shown to correlate with other morphological parameters, such as the concentration index. Likewise, $M_{20}$, the second order moment of the 20\% brightest pixels in the galaxy, and the asymmetry, $A$, which depends on the residuals of an image after subtracting a 180 degree rotation from itself are used to morphologically classify galaxies and identify mergers.  

In the following analysis we apply these nonparametric measures to the PSF-subtracted images (second column of Figure \ref{fig:hst_images}) for the blue and red filters separately. 
To mitigate the effects of residual flux from PSF-subtraction, we exclude the flux from a 4-pixel-radius circle around the quasar's position.  

We chose not to consider two other nonparametric quantities often used to asses the degree of merging in galaxies: the concentration parameter, $C$, assesses the flux distribution in a source by the ratio of flux in a circle (ellipse) with a small radius (semimajor axis) to a larger one encompassing most of the flux; and the smoothness parameter, $S$, measures clumpiness in a galaxy by subtracting a smoothed image from the raw to quantify small-scale fluctuations \citep[together with $A$ comprise the ``CAS" system;]{Conselice03}.  The concentration index depends on inner galaxy light which we cannot accurately assess; the Gini coefficient does a better job of measuring concentration for our sources. The smoothness parameter loses efficacy at high redshift where a resolution element (${\rm PSF}_{\rm FWHM} \sim 0.15 - 0.2$\arcsec) corresponds to $\sim 1.5$ kpc.  Therefore, in this paper, we concentrate on $G$, $M_{20}$, and $A$ to study the morphologies of red quasar host galaxies.  

Since the depths of our 22 images vary and since the redshift of our sample introduces surface brightness dimming as $(1+z)^4$, we follow the technique outlined in \citet{Lotz04} to generate segmentation maps that define a set of pixels to analyze that are above a uniform threshold for the entire sample.  We compute the surface brightness per pixel, as a function of radius (where the center of the map is the peak position of the quasar), $\mu(r)$.  We compute the Petrosian radius, $r_p$, defined as the radius at which the surface brightness is 20\% the mean surface brightness within that radius, i.e., 
\begin{equation}
\eta = \frac{\mu(r_p)}{\bar{\mu}(r<r_p)} \ {\rm with}\ \eta = 0.2. \label{eqn:r_p}
\end{equation}
The segmentation map contains all the pixels with $\mu > \bar{\mu}(r_p)$.
 
We then create an array, $X_i$, sorted in order of increasing pixel value, and compute the Gini coefficient using the algorithm:
\begin{equation}
G = \frac{1}{\bar{X}n(n-1)}\sum_i^n (2i-n-1)X_i, \label{eqn:gini}
\end{equation}
where $n$ is the total number of pixels in the segmentation map and $\bar{X}$ is the mean of all $X_i$ values.  

\begin{figure*}
\plottwo{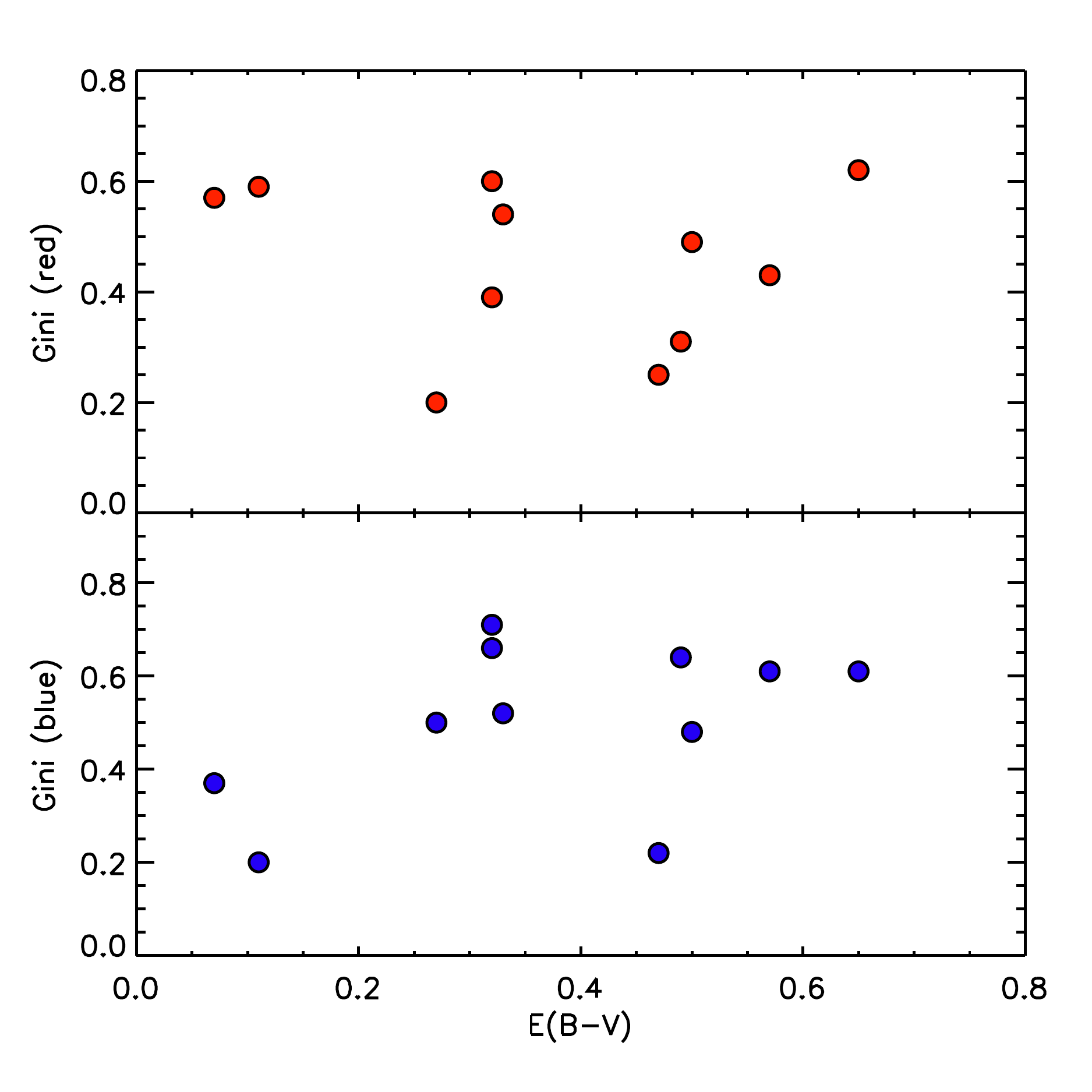}{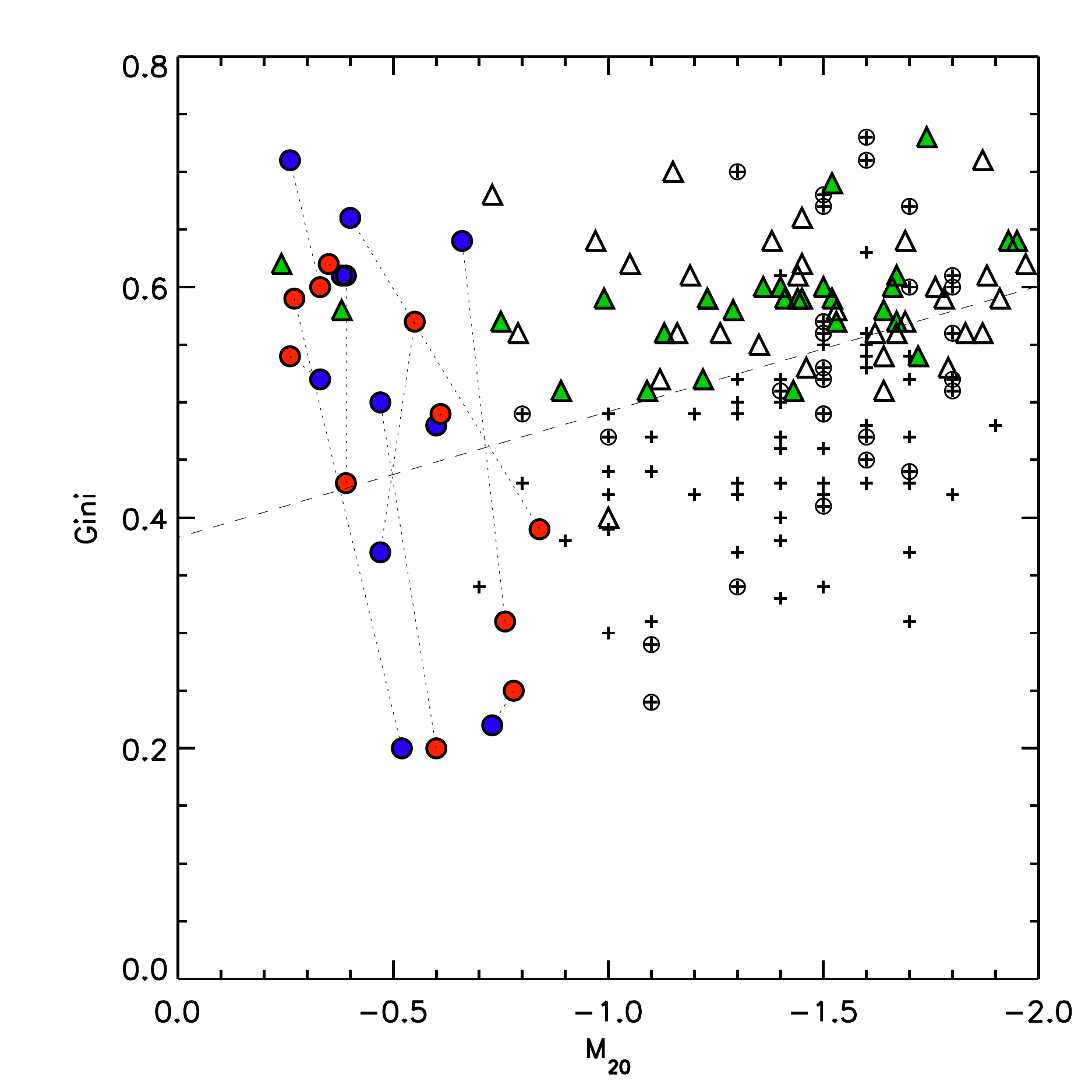}
\caption{{\em Left --} Gini coefficient versus $E(B-V)$ measured in the red filter (top panel) and the blue filter (bottom panel).  We see no correlation between the two quantities.  
{\em Right --} $M_{20}$ versus Gini coefficient for the red quasar host galaxies measured in the blue and red filters, and color coded accordingly.  The two measurements for each object are connected by a dotted line. Plotted for comparison with crosses are the same quantities computed for dusty star forming galaxies at $z\sim2$ from CANDELS \citep{Wang12} and the circled crosses are X-ray sources in that sample.  Local ($z\le0.24$) ULIRGs from \citet{Borne00} are plotted with triangles and ULIRGs identified as having ``Double'' nuclei are colored green.  The dashed line is the separation between ``normal'' and disrupted galaxies defined by \citet{Lotz04}.  Our red quasar hosts are consistent with the same Gini values as the comparison samples, but only the most extreme ULIRGs with ``double'' nuclei have $M_{20}$ values consistent with the red quasars.}
\label{fig:m20_gini}
\end{figure*}

\citet{Urrutia08} found a correlation between the Gini coefficient and $E(B-V)$, indicating that redder sources are more disturbed.  
The same plot for our $z\sim2$ red quasars (Figure \ref{fig:m20_gini}, left) shows large scatter and no such correlation. 

The Gini coefficient alone does not determine the degree of merging/disruption, since a de Vaucouleurs profile ($n=4$) is more centrally concentrated than a disk, yet is still a smooth light distribution.  \citet{Lee13} find that passive, elliptical, CANDELS galaxies at $z\sim 2$ have $G$ between 0.4 and 0.7, with a higher mean than the star forming systems (0.53 versus 0.43) largely due to their light profile being concentrated in a single central peak.  However, a merger with bright clumps of star formation will also have large $G$ values, making this parameter by itself insufficient for identifying merging systems.  

The distinction between a centrally concentrated light profile and a clumpy merger can be made when $G$ is combined with a second parameter, such as $M_{20}$, which represents the second order moment of the $20\%$ brightest pixels in a galaxy's light profile \citep{Lotz04}.  $M_{20}$ is defined as follows:  
\begin{equation}
M_{20} = \log10\Bigg( \frac{\sum_i M_i}{M_{\rm tot}}\Bigg) \label{eqn:m120}
\end{equation}
where 
\begin{equation}
M_{tot} = \sum^n_i M_i = \sum_i^n X_i r_i^2.
\end{equation}
Here $n$ and $X_i$ are the total number of pixels in the segmentation map and the flux per pixel, as defined for equation \ref{eqn:gini}.  We sum over $M_i$ while $\sum_i X_i < 0.2X_{\rm tot}$.  \citet{Lotz04} showed that smooth light profiles correspond to low values of $M_{20}$, while extended sources with clumps or multiple nuclei (i.e., mergers) have high values, with $M_{20} \ge -1.1$.  In general, the further the brightest pixels are from the center of the source, the closer $M_{20}$ is to a value of zero.

The right-hand panel of Figure \ref{fig:m20_gini} shows the relationship between $M_{20}$ and the Gini coefficient for both the red and blue filters.  For comparison we plot with crosses CANDELS galaxies selected to have $1.5 \le z \le 2.5$ and $M_{\star}=10^{11} M_\odot$ and morphologically analyzed by \citet{Wang12}.  To compare our sample to morphologically similar galaxies, we also plot with triangles a sample of 73 local ($z\le 0.24$) ULIRGS studied with {\em HST} in the rest frame optical by \citet{Borne00}.  These sources were used by \citet{Lotz04} to compare with their sample of $z\sim2$ galaxies observed with {\em NICMOS}, making them a suitable comparison set to our sources as well.  
Their morphological properties, including $G$, $M_{20}$ and $A$ as well as a multiplicity classification, are presented in Table 5 of \citet{Lotz04}.  The sources with ``Double'' nuclei are colored green.  

The dashed line indicates the separation between ``normal'' and disrupted galaxies, as defined in \citet{Lotz04}.  All but one of our red quasars reside above the line in at least one of the filters.  
The local ULIRGs, particularly those with a ``Double'' nucleus morphology, have the lowest  $M_{20}$ values and most closely approach the red quasar sample.

However, the $M_{20}$ values of the red quasar sample are systematically shifted toward higher values with respect to the ULIRGs.  Only the most extreme ULIRGs with ``double'' nuclei are consistent with the red quasar sample.  The higher $M_{20}$ values of the red quasar sample indicate host galaxy light distributions where the brightest pixels tend to be farther away from the central nucleus.  We note that it is conventional to plot $M_{20}$ with lower values to the right, so the red quasars have the highest $M_{20}$ values and appear to the left in Figure \ref{fig:m20_gini}.

\begin{figure*}
\plotone{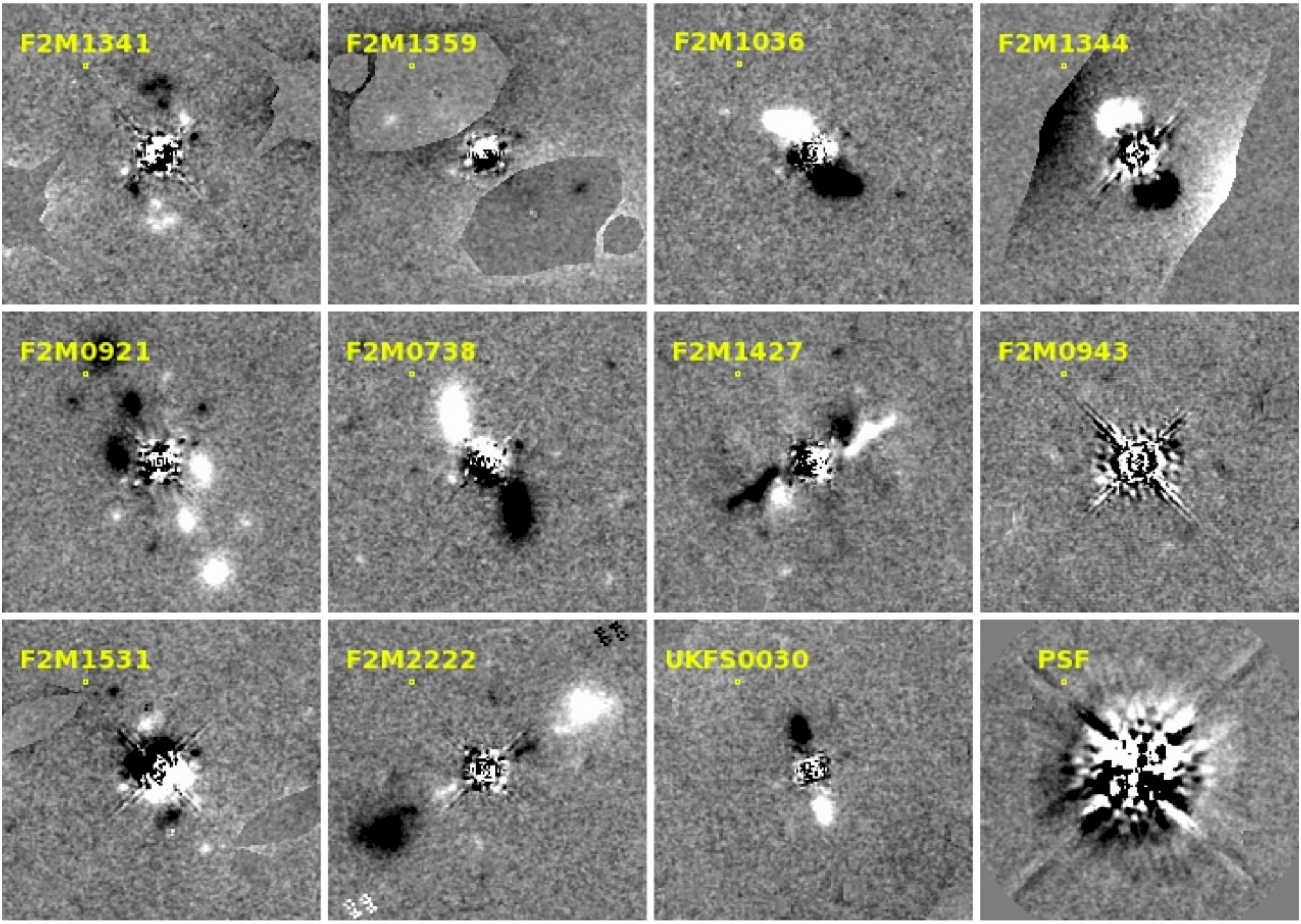}
\caption{Asymmetry maps for the F160W images produced by rotating the masked PSF-subtracted image, $I$, by $180^\circ$ to produce the rotated image, $I_{180}$, and taking the difference ($I-I_{180}$). We use these maps in the numerator of equation \ref{eqn:asymm} to compute $A$. The bottom right panel shows the asymmetry of the F160W PSF.  These maps clearly show the complex and detailed structure surrounding the residual of the point source in most of our sources; the exceptions are F2M1359 and F2M0943.  }
\label{fig:asymm}
\end{figure*}

The third metric that has been shown to effectively separate normal from merging galaxies is the rotational asymmetry, $A$.  The standard prescription for computing $A$ involves rotating the image of a source, $I$, by 180$^\circ$ (creating the rotated image $I_{180}$) and producing asymmetry maps by taking the difference between $I_{180}$ and $I$.  The asymmetry maps are then used in the following formalism to quantify asymmetry:
\begin{equation}
A = \frac{\sum_{i,j} | I(i,j) - I_{180}(i,j) |}{\sum_{i,j} |I(i,j)|}. \label{eqn:asymm}
\end{equation}
Since our sources are dominated by a strong point source at the center, we experimented with rotating and subtracting the original image, as well as the PSF-subtracted frame (second column of Figure \ref{fig:hst_images}) and found that the latter produces cleaner PSF-removed asymmetry maps.  Figure \ref{fig:asymm} shows the asymmetry images, $I-I_{180}$ for all eleven quasars as well as the PSF for the F160W images.  In addition to clearly exposing the underlying disrupted host galaxies for most of the sources, we can re-examine three sources previously deemed undisturbed.  F2M0943 continues to show no underlying structure, F2M2222 exposes a small jet-like protrusion near its core that was hidden by the residual PSF light in the second and fourth panels of Figure \ref{fig:hst_images}.  
F2M1359, which we characterize as being serendipitously reddened by the intervening galaxy, shows no asymmetry in its image, consistent with its host having a smooth symmetric profile as found by Galfit. 

We compute $A$ for the 12 red quasars following equation \ref{eqn:asymm} using the images shown in Figure \ref{fig:asymm}.  To avoid PSF-related effects to affect our computation we exclude the innermost $16\times16$ pixels from our analysis and include only unmasked regions.  The asymmetries we find have very high values ($A = 0.99 - 1.6$), significantly higher than those found for local ULIRGs in \citet{Lotz04}.  However, we also found that the calculation of $A$ is extremely sensitive to how the background is defined as well as whether segmentation maps are used, versus the full image.  

Collectively, the Gini coefficient, $M_{20}$ and $A$ all suggest a high degree of asymmetry and clumpiness, as expected from merger-induced tidal effects and star formation.
Table \ref{tab:nonparam} lists the Petrosian radii, Gini coefficients and $M_{20}$ computed for the two filters.

\section{Discussion of Individual Quasars}\label{sec:indiv_qsos}

In this section, we discuss each source individually in increasing redshift order, as they appear in Figures \ref{fig:hst_images} and \ref{fig:hst_images2} and note unique aspects of the morphological fits and host galaxy properties that we can infer.  Throughout this section, statements about the ``projected distance" to a companion system implicitly assume that the companion is at the redshift of the quasar.  

\subsection{F2M1341}\label{sec:f2m1341}
The PSF-subtracted residual frame in Figure \ref{fig:hst_images} shows smooth, arc-like emission above and below the central point source.  These are fit by two \sersic\ components in both filters.  The southern component has a disk-like \sersic\ index of $n_{\rm F105W}=1.3$ with an effective radius of $R_e = 6.4$ kpc at a projected distance of 11.4 kpc.  The redder component's \sersic\ index and effective radius both have larger fitted values but are highly uncertain.  The projected distance of the red component is a more distant 12.4 kpc.

More intriguing is the northern component which has well-separated blue and red emission concentrations.  The blue component is nearer to the quasar, and has a de Vaucouleurs-like profile with $n_{\rm F105W}=4.8$ while the red component is fit by a shallow, Gaussian or disk-like profile with $n_{\rm F160W} = 0.64$.  The two components are separated by $\sim 4$ kpc.  In addition, there are two faint, red point-like sources to the east which we did not include in the fit as they may be faint low-mass Galactic stars. 

\subsection{F2M1359}\label{sec:f2m1359}
This source has a host galaxy component detected in both F105W and F160W bands whose S\'{e}rsic indices indicate a significant bulge component, although the precise value of $n$ is uncertain ($n_{\rm F105W} = 2.54\pm0.4$, $n_{\rm F160W} = 8.0\pm4.9$).  Since this source is the only object with a unambiguous detection of a single host galaxy component at the quasar's position with no additional components, the most likely scenario is one in which there is no merger.  
Additionally, there is a large, lower-redshift galaxy to the north of the quasar whose \sersic\ index indicates that it is disk dominated ($n=1.35$), suggesting that the reddening in this case is due to extinction from the extended disk of the larger nearby galaxy.  

This would mean F2M1359 is an accidental red quasar.  With an extinction corrected absolute-$K$-band magnitude of $-29.7$ (AB) it has the eighth highest luminosity of our sample, which is still remarkably luminous as compared with the unreddened quasars (black points in Figure \ref{fig:absKz}).  
Among the 13 red quasars studied by \citet{Urrutia08}, one source (F2M0834+3506) was found to also be a normal quasar reddened by an intervening galaxy, so statistically it appears that $\lesssim 10\%$ of red quasars may be due to reddening that is not intrinsic to the quasars' host galaxies.

\subsection{F2M1036}\label{sec:f2m1036}

A highly asymmetric system with strikingly separated blue and red components that are best fit by two separate \sersic\ components in each band.  All the components have small \sersic\ indices ($n<1$).  Some red light is seen in the residual image north-east of the source, which is not fully captured by the fitting routine and is likely responsible for the higher $\chi^2$ value in the F160W image (6.00 versus 2.92). 

\subsection{F2M1344}\label{sec:f2m1344}

\begin{figure}
\epsscale{1.2}
\plotone{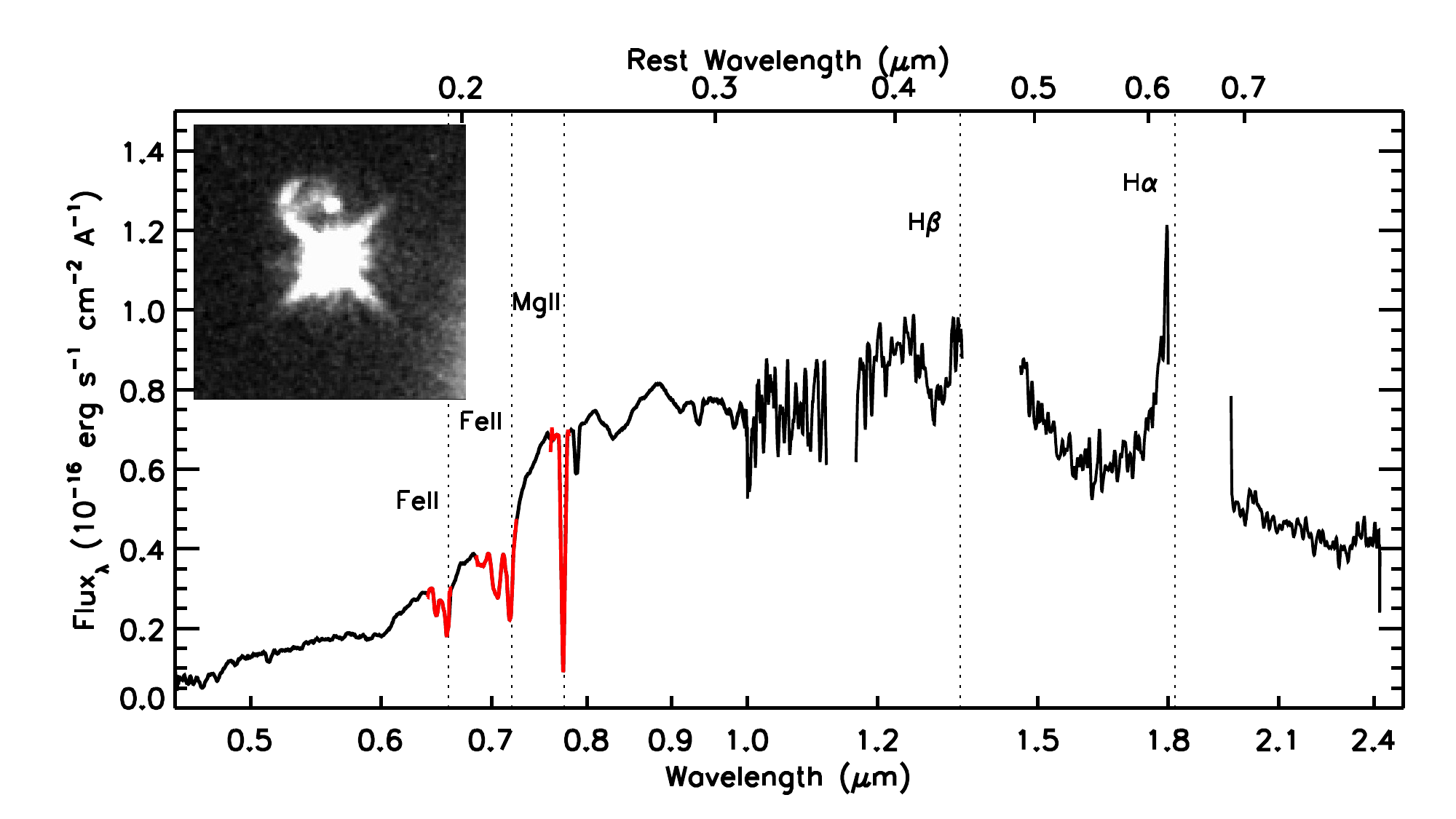}
\caption{Optical through near-infrared spectrum of F2M1344+2839 demonstrating its unusual spectral shape and classification as a FeLoBAL by \citet{Urrutia09}.  The broad absorption features in \ion{Mg}{2} and \ion{Fe}{2} are marked with a red line and extend for 2000-5000 km s$^{-1}$.  The inset shows a $6\arcsec\times6\arcsec$ image in the F105W filter showing extended emission suggestive of tidal tails or a disrupted companion galaxy.  There is also a faint point source visible in the frame.  }
\label{fig:f2m1344}
\end{figure}

Despite being near a large intervening galaxy, unlike F2M1359 (\S \ref{sec:f2m1359}) this source shows independent evidence for a merging host galaxy.   The multi-component residual seen in the PSF-subtracted frame (second column) to the north of the PSF is, on its own, suggestive of a merger. 
There is also a faint point-like source in the model for the companion galaxy, which can be better seen in the inset of Figure \ref{fig:f2m1344} that shows a $6\arcsec\times6\arcsec$ image of this quasar in the F105W filter.  This source was better fit by a \sersic\ profile with the large and unphysical indices $n_{\rm F105W} = 20.00$ and $n_{\rm F160W} = 7.17$.  Such a profile is centrally concentrated and indicates that this source may be a second AGN or perhaps a luminous, compact clump of star formation.

The combined optical plus infrared spectrum of the quasar shows complicated absorption, including blue-shifted absorption in \ion{Mg}{2} and \ion{Fe}{2} \citep{Urrutia09,Glikman12}, as seen in the optical-to-near-infrared spectrum shown in Figure \ref{fig:f2m1344}. While some of the UV absorption may be due to dust in the intervening galaxy, there are clearly in situ absorbing systems indicative of a merger. 

\subsection{F2M0921}\label{sec:f2m0921}

\begin{figure}
\epsscale{1.2}
\plotone{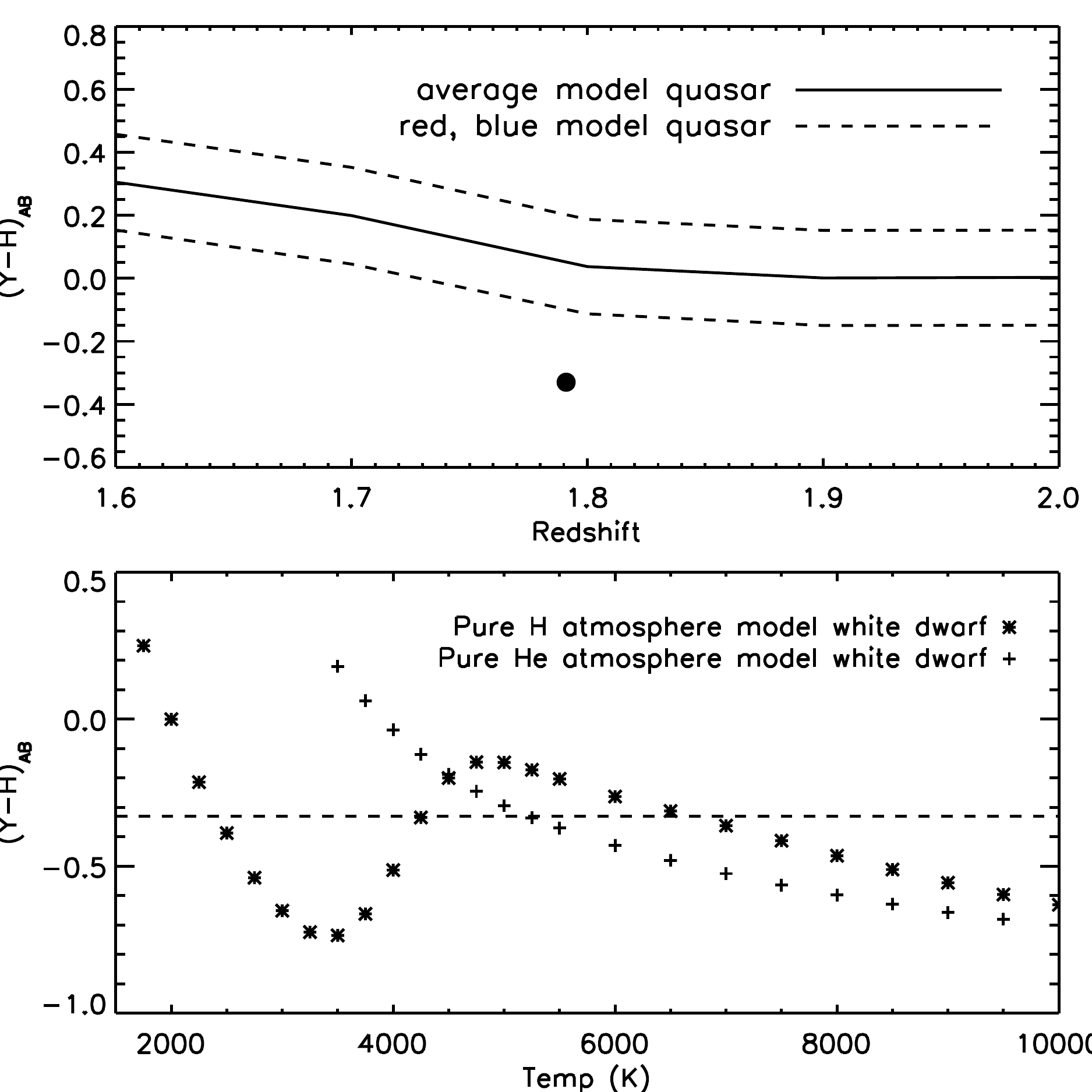}
\caption{Predicted $Y-H$ colors of quasars (top panel) and white dwarfs (bottom panel) from the UKIDSS colors derived by \citet{Hewett06} are plotted to compare with the point source seen $\sim 4$\arcsec\ away from red quasar F2M0921 (black circle in the top panel; horizontal line in the bottom panel).  The color of this source is too blue to be a quasar at this redshift, but has colors consistent with a Galactic white dwarf. }
\label{fig:f2m0921}
\end{figure}

This source has one of the clearest indications merging galaxies offset from the position of the quasar.  There is also a point source 4\arcsec\ away that we fit in this model.  
From the Galfit photometry, the color of this second point source in the two filters is $-0.11$.  When we correct this color for the offsets to the UKIDSS passbands that were derived in Section \ref{sec:obs}, the point source has $Y-H=-0.33$.  
The top panel of Figure \ref{fig:f2m0921} plots the $Y-H$ color of our source compared to colors of quasars as a function of redshift, based on the synthetic UKIDSS colors derived for quasars \citep[Tables $25-27$ of][]{Hewett06}.  The filled circle is the source's Galfit colors.  
At $z\sim1.8$ (the redshift of this source is $z=1.791$) the $Y-H$ color ranges between $-0.1$ and $0.2$ mags. 
The point source source is significantly bluer than a typical unobscured quasar at this redshift.  We therefore rule out this source as a companion quasar, which would be at a projected distance of 35 kpc away.  

The bottom panel of Figure \ref{fig:f2m0921} shows the $Y-H$ color versus temperature for two white dwarf models,  from Tables 13 and 14 \citet{Hewett06}.  The colors of this source are consistent with a white dwarf.  And using the absolute magnitudes corresponding to the temperatures best-agreeing with this source's color places such a white dwarf between $\sim 40$ and 330 pc.  

\subsection{F2M0738}\label{sec:f2m0738}

This source is $\sim 0.5$\arcsec away from a companion galaxy,  which -- if at the same redshift as the quasar -- is at a projected distance of 16.4 kpc from the quasar.  The galaxy is well fit by reasonable \sersic\ indices ($n_{\rm F105W} = 2.50$ and $n_{\rm F160W} = 3.73$) and corresponding effective radii of 4.3 kpc and 3.2 kpc, respectively. We also fit an underlying host with \sersic\ indices of $n_{\rm F105W} = 1.85$ and $n_{\rm F160W} = 2.77$, consistent with a bulge plus disk hybrid.  

\subsection{F2M1427}\label{sec:f2m1427}

This system has one of the most complex morphologies of our sample.  The raw image shows a protruding structure to the south-east side of the quasar, and the PSF-subtracted frame in Figure \ref{fig:hst_images} shows complex structure, extended diffuse emission and $\sim 5$ red point sources. The radio contours are circular and symmetrically centered on the optical peak with no evidence of elongation lined up with the optical feature.  However, the FIRST beam has a FWHM of 5\arcsec with 1\farcs8 pixels, and higher resolution radio images may reveal more complex structure.  

We model the significant emission by three \sersic\ components which capture much of the flux, but are likely not physically representative of the host systems.  We conclude this because the F105W and F160W fit parameters do not agree well with each other.   This is most likely another example of a very complex, multi-component merger similar to the systems seen in \citet{Urrutia08}, but with a loss of the low surface brightness features needed to fully reconstruct the details of system.  

\subsection{F2M0943}\label{sec:f2m0943}

This source would not converge with any physically meaningful added \sersic\ component.  A single PSF component yielded a fit with reduced $\chi^2$ values of 51.7 and 73.8 in F105W and F160W, respectively.  When a \sersic\ component is added, the reduced $\chi^2$ improves greatly (though still leaving behind a strong residual) to 15.6 and 23.1 in the two filters, respectively, but the effective radii for the \sersic\ components are unphysical, at 0.01 and 0.03 pixels, respectively\footnote{Adding a second PSF rather than a \sersic\ component performs more poorly, yielding reduced $\chi^2$ values of 21.2 and 24.9 in the blue and red filters, respectively.}. In addition, Galfit assigns the added \sersic\ component more flux than the PSF component (i.e., the \sersic\ component is brighter than the PSF component by $0.7-0.8$ magnitudes).  

We consider possible explanations for this poor fitting outcome.  One possibility is that this source, because of its higher redshift, was observed over two orbits. Breathing of the telescope or imperfect image combining via {\tt astrodrizzle} may have affected the shape of the PSF in this field.  We investigated the shapes of the three stars from the full reduced imaged of F2M0943 that went into creating our master PSF in both filters and did not find significant systematic offsets in the FWHM or other shape parameters as compared with stars in our other fields.  We also constructed `mini-PSFs'  out of the stars in each field to look for evidence of differences between this field's PSF compared with the others, and did not find any differences.  The asymmetry map for this source also shows the strongest residuals in the point source, and does not show an obvious underlying host.  

Ruling out large PSF variations, we interpret the problematic nature of fitting this source as either (a) there is something intrinsically different about this source (i.e., it is a luminous quasar hosted by an extremely luminous and centrally concentrated host), or (b) this object, being the most luminous quasar in our sample, is so bright that the Poisson noise dominates the residuals in the PSF fitting.  Given that the residual image has low asymmetry and that the interpretation in (a) is highly unphysical, we propose that (b) is the more likely explanation.  In  support of this interpretation, we note that this source has the highest dynamic range (as defined in Section \ref{sec:results}) and that residuals of this significance are seen in some of the most luminous sources in \citet{Floyd04}.

\subsection{F2M2222}\label{sec:f2m2222}

This source is well-fit by a point source with the only nearby galaxy seen at a projected distance of 32 kpc.  The nearby galaxy is a blue, disk-dominated ($n_{\rm F125W} = 1.62$ and $n_{\rm F160W} = 1.64$) galaxy with an effective radius of $R_e = 6.1$ kpc and $R_e = 5.8$ kpc in the F125W and F160W filters, respectively, assuming it is at the same redshift.  The residual image shows excess blue clumps in this galaxy suggestive of enhanced star formation.  However, it is not clear whether this galaxy is physically associated with the quasar and its large projected distance indicates that this quasar is likely not hosted by an obvious merger.  

\subsection{F2M1531}\label{sec:f2m1531}

This object is well-modeled by a central source plus a host galaxy that is 2.0 and 1.1 magnitudes fainter in F125W and F160W, respectively, and has a \sersic\ index indicative of a bulge ($n_{\rm F125W} = 3.64$ and $n_{\rm F160W} = 4.48$). However, Figure \ref{fig:hst_images2} shows that some residual PSF flux remains in the image after the point source is subtracted. While there is a clear detection of an extended host galaxy, we consider its fitted flux to be an upper limit.

Additionally, an elongated feature to the south is fit by a Gaussian profile ($n_{\rm F125W} = 0.67$ and $n_{\rm F160W} = 0.61$) at a projected distance of 12.3 kpc.  This may be a tidal tail or the core of an interacting companion galaxy.  

\subsection{UKFS0030}\label{sec:ukfs0030}

This source is the lowest-luminosity and highest-redshift source in this sample (de-reddened absolute $K$-band magnitude of $-30.99$ at $z=2.299$).  It is the only red quasar observed from the UKFS sample of \citet{Glikman13}.  This source is well-fit by a \sersic\ component at the quasar location (with a central position at a projected distance of 0.9 kpc from the quasar) but only in the F125W filter.  The component parameters are physically consistent with a concentrated bulge/disk hybrid ($n_{\rm F105W} = 2.74$) and $R_e = 1.5$ kpc, which may be interpreted as a nuclear starburst.  There is also a nearby component to the northwest, at a projected distance of $\sim 11$kpc.  

\section{Merger Statistics for Red Quasars}\label{sec:disc}

Using a wide range of surveys from the literature for which AGN host morphologies are studied, \citet{Treister12} examined the merger fraction among AGNs as a function of luminosity and redshift and found a strong dependence on luminosity over three orders of magnitude.  The merger fractions ranged from 4\% in {\em Swift}/BAT-detected AGN \citep[largely low redshift systems at $z<0.05$, ][]{Koss11} with $L_{\rm bol} \sim 10^{43.5}$ erg s$^{-1}$, up to 85\% for the red quasars studied by \citet{Urrutia08}, whose bolometric luminosities were the highest in the studied sample ($L_{\rm bol} \sim10^{46.2}$ erg s$^{-1}$).  

Here we place the $z\sim2$ red quasars on the merger fraction vs.~luminosity plot shown in Figure 1 of \citet{Treister12}.
Out of the eleven quasars studied in this paper, at most three (F2M1359, F2M0943, F2M2222) sources do not show evidence for a galaxy merger.  
However, we concluded that F2M1359 is not an intrinsically dust-reddened quasar and is likely obscured by an intervening galaxy lying along the line of sight, reducing the denominator in our calculation to ten intrinsically reddened quasars.   F2M2222 is a more ambiguous case, leaving  just one quasar (F2M0943) as showing no evidence at all for a merger. 
Therefore, we conservatively compute a merger fraction of 8/10 or 80\%, although it could be as high as 90\%.

To compute the bolometric luminosities of the $z\sim 2$ red quasars, we use bolometric correction from the quasar SED of \citet{Richards06}.  Because of reddening, we cannot use the optical or near-infrared magnitudes that we have presented throughout this work.  For the thirteen red quasars in \citet{Urrutia12}, full SED modeling was performed including {\em Spitzer} IRS spectra and MIPS photometry out to 160 $\mu$m, to compute their bolometric luminosities.  Without data spanning such a broad wavelength range, we estimate the bolometric luminosities of our quasars by matching them to the Wide-Field Infrared Space Explorer \citep[WISE;][]{Wright10} all-sky source catalog.  All of our sources are detected within 1\arcsec\ in all four WISE bands. We use the longest-wavelength band, $W4$, whose effective wavelength is 22.0883 $\mu$m (corresponding to rest-frame wavelength between 6.1 $\mu$m and 8.1 $\mu$m), in order to minimize the effects of dust extinction and probe the intrinsic quasar emission.  At these wavelengths, the bolometric corrections from the \citet{Richards06} SED\footnote{We use the SED that is made up of all the SDSS quasars in that sample.} are all a factor of $\sim 8$.  Using this method, our quasars' luminosities have a range $\log(L_{\rm bol}) = 47.8 - 48.3$ (erg s$^{-1}$). 

\begin{figure}
\plotone{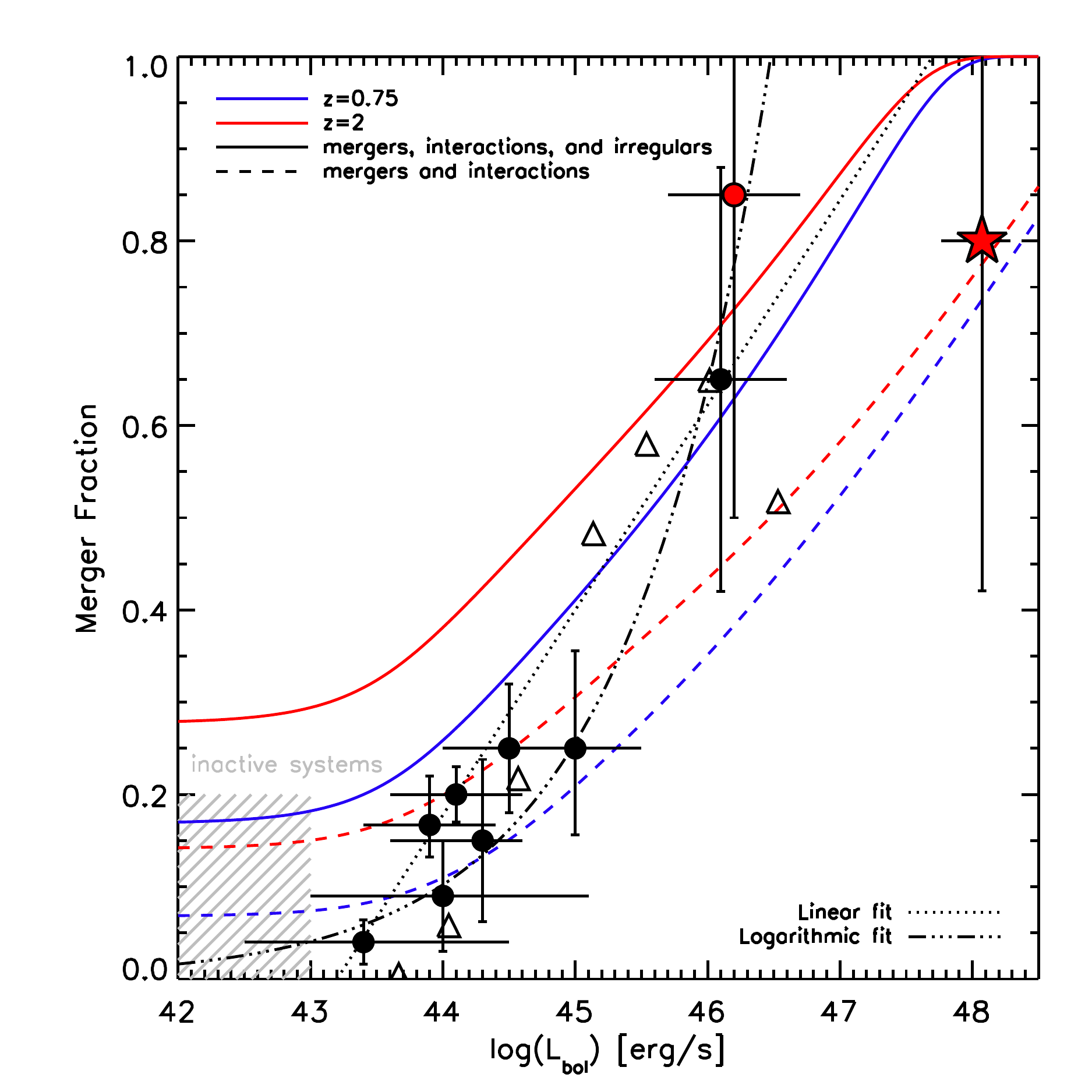}
\caption{Merger fraction as a function of bolometric luminosity using the data from Table 1 of \citet{Treister12} (black points). The red star represents the sources in this work.  The dotted and dot-dot-dot-dashed black lines represent the linear and logarithmic fits to the points presented in \citet{Treister12}.  Our data have effectively as high a merger fraction (80\%) as the next highest point (red circle; 85\%) of \citet{Urrutia08} but are more than an order of magnitude more luminous.  The blue and red lines show the predicted merger fraction based on the variability-based model from \citet{Hickox14} realized at $z=0.75$ and $z=2$.  The solid line includes ``mergers, interactions, and irregulars'' in determining the fraction, while the dashed line only considers ``mergers and interactions'', which effectively ignores minor mergers.  The merger fraction in our sample agrees well with the ``mergers and interactions'' model at $z=2$ and suggests that major mergers play the most significant role in fueling the most luminous quasars at $z=2$.}
\label{fig:fmerge}
\end{figure}

Figure \ref{fig:fmerge} shows the merger fraction in AGN samples across many orders of magnitude in AGN luminosity.  Circles and triangles are data from a variety of AGN host galaxy studies and was incorporated into a meta-analysis of AGN triggering mechanisms by \citet{Treister12}. This study of merging red quasar hosts at $z\sim2$ adds the most luminous AGN sample to this plot (red star). Our $z\sim2$ red quasars are significantly more luminous but their merger fraction is commensurate with the 85\% found at $z=0.7$ (red circle). The dotted and dot-dot-dot-dashed lines are the parametrized linear and logarithmic fits to the data \citep[equations 1 and 2 of ][]{Treister12}.  

For a physical interpretation, we compare our result to the variability-driven model of \citet{Hickox14}.  Blue and red lines represent the predictions for merger fraction of AGN at $z=0.75$ \citep[the][sample]{Urrutia08} and $z=2$ (this work), respectively.  While the error bars on our data point are too large to favor any particular model definitively, we note that while the lower-luminosity points agree better with the model that includes ``mergers, interactions, and irregulars" (solid line), at the luminosities of our sources, the merger fraction is in better agreement with the model that only includes ``mergers and interactions'' (dashed line).  The latter model effectively excludes the role of minor mergers, which is a reasonable consideration at the luminosity regime of our sample.  Therefore, it is likely that the role of minor mergers in black hole accretion declines with luminosity and is yet another lever that must be calibrated in simulations of cosmic AGN fueling.  

With a surface brightness limit of $\sim24$ mag arcsec$^{-2}$ (Figure \ref{fig:profile}) the depth of our imaging is sensitive to major mergers but not sensitive to minor mergers.  Thus the merger fraction we report in this paper is effectively that due to major mergers only.  The fraction is consistent with the predicted contribution of major mergers hosting the most luminous AGN according to \citet[][dashed lines in our Figure \ref{fig:fmerge}]{Hickox14}.  The solid lines in Figure \ref{fig:fmerge}, which include minor mergers as predicted by \citeauthor{Hickox14}, suggest that were our images deep enough to detect minor mergers, 100\% of the sample would show evidence of minor or major merging activity.

However, we caution that the quasars studied by \citet{McLure99},  \citet{Kukula01}, \citet{Dunlop03} and \citet{Floyd04} were also of blue, luminous quasars with $-24 \le M_V \le -25$, yet found that their hosts are largely passive elliptical galaxies and they find no evidence for merging hosts.  The most luminous blue quasars presented by \citet{Floyd04}, reaching $M_V \lesssim -28$ at $z<0.5$, also have elliptical profiles with only one (out of 17) objects showing clear signs of a merger. This is consistent with the idea that the quasars presented here and in \citet{Urrutia08} were selected to be {\em reddened} Type I (i.e., showing broad emission lines) and are therefore in an earlier stage of the merger-driven evolutionary sequence, before the host galaxies' morphologies settle into a virialized elliptical galaxy.  

This implies that even more heavily reddened quasars than the F2M and UKFS samples should show even stronger signs of mergers. Such sources may be hard to detect, but could be found by selecting a sample of X-ray sources with comparable redshifts and luminosities with no optical counterpart or spectroscopic evidence for AGN activity.  Since such sources are expected to be rare, a large-area survey would be needed.  Such surveys are now becoming available, e.g., Stripe 82X \citep{LaMassa13a,LaMassa13b} and the XMM-LSS surveys \citep{Pierre07,Chiappetti13}.  

Interestingly, the morphologies of our quasars suggest that they are in an early stage of the mergers, with clear independent companions rather than asymmetric or disrupted single hosts.   However, the models \citep[e.g.,][]{Hopkins05c,Hopkins06a,Hopkins08} predict that the AGN are brightest for longest in the late merger stage. One possible explanation is that these red quasars are associated with the early merger stages when the dust is typically far from the nucleus, rather than the late-stage merger when the dust can be more concentrated around the nucleus \citep[as seen in the heavily obscured Type II quasars in ][which seem to be mostly late-stage mergers]{Lacy07b}.  Another interpretation is that the red quasars are tracing group interactions in which the quasar host has already been through a merger, but additional mergers are ongoing.

\section{Summary and Conclusion}\label{sec:summ}

We have conducted a morphological study with the WFC2/IR camera on {\em HST} of eleven $z\sim 2$ dust reddened quasars that are intrinsically the most luminous systems at this redshift.  This is the highest redshift sample of {\em HST}-imaged dust-reddened quasars. Visual inspection shows clear evidence of mergers in at least eight of the eleven systems; only one source does not appear to reside in a merging host.

We performed careful PSF modeling in the three broad near-infrared filters in order to remove the quasar light and study the morphologies of their underlying host galaxies.  We modeled the galaxies with parametric fitting to a \sersic\ profile and see significant merging components in eight systems.  One system proved to be reddened by an intervening galaxy and is not technically a ``red quasar''.  We use the \sersic\ profiles to integrate the total flux of the host galaxies to study their properties and find that these galaxies are more luminous than the massive galaxies at $z\sim 2$ studied by the CANDELS survey.
Although redder than the overall systems' colors, the red quasar host galaxies are mostly bluer than the CANDELS galaxies.  Three sources are significantly redder, suggesting a very dusty star bursting host.  We interpret the diversity of colors coupled with high luminosities as consistent with these systems still exhibiting ULIRG-like properties as a result of a major merger.  

We conduct a non-parametric analysis of the PSF-subtracted images, measuring their Gini parameter, $M_{20}$ and asymmetry and find broad consistency with local ULIRG properties,  particularly with the most extreme double-nuclei ULIRGs, further supporting them being hosted by major mergers.  

Compared with studies of the merger fraction in AGN at different luminosities, our sample is more luminous by two orders of magnitude than the previous most-luminous sample of red quasars studied in this manner \citep{Urrutia08} and has a comparable merger fraction.   When added to AGN samples from other studies at a variety of AGN luminosities on a plot of host galaxy merger fraction versus AGN luminosity \citep{Treister12} and compared with variability-driven black hole growth models \citep{Hickox14}, our data favors a model in which black hole growth in the most luminous AGN at $z \sim 2$ is dominated by major mergers.
We conclude from this study that major mergers are the dominant drivers of black hole growth at the highest luminosities as far back as the epoch of peak quasar and star formation activity.





\acknowledgments

%
TOPCAT \citep{taylor05} and an OS X widget form of the JavaScript Cosmology Calculator \citep{wright06,simpson13} were used while preparing this paper. \\
We thank Ezequiel Treister for sharing data to re-create Figure \ref{fig:fmerge}, Ryan Hickox for providing the simulated merger fraction curves in Figure \ref{fig:fmerge}, and the anonymous referee for helpful comments that improved this manuscript. The authors gratefully acknowledge Space Telescope Science Institute for support via Hubble grant GO-12942. Support for program \#12942 was provided by NASA through a grant from the Space Telescope Science Institute, which is operated by the Association of Universities for Research in Astronomy, Inc., under NASA contract NAS 5-26555. EG acknowledges the generous support of the Cottrell College Award through the Research Corporation for Science Advancement.  
BDS gratefully acknowledges support from the Oxford Martin School, Worcester College and Balliol College, Oxford. KS gratefully acknowledges support from Swiss National Science Foundation Grant PP00P2\_138979/1. CMU thanks Yale University for support. 

The National Radio Astronomy Observatory is a facility of the National Science Foundation operated under cooperative agreement by Associated Universities, Inc.

This work uses observations taken by the CANDELS Multi-Cycle Treasury Program with the NASA/ESA HST, which is operated by the Association of Universities for Research in Astronomy, Inc., under NASA contract NAS5-26555.
{\it Facilities:} \facility{HST}, \facility{Palomar}, \facility{IRTF}.

\bibliography{hst_rq_astroph.bbl}




\begin{deluxetable*}{cccccccccccccc}


\tabletypesize{\footnotesize}

\tablewidth{0pt}

\tablecaption{Table 1: High Redshift Red Quasars Imaged with {\em HST WFC3/IR} \label{tab:qso_sample}}


\tablehead{\colhead{Name} & \colhead{R.A. } & \colhead{Decl.} & \colhead{$J$} & \colhead{$K_s$} & \colhead{$A_K$} & \colhead{Redshift} & \colhead{Orb} & \colhead{Filter} & \colhead{Exp} &  \colhead{$\mu$}\tablenotemark{a} & \colhead{Filter} & \colhead{Exp} & \colhead{$\mu$}\tablenotemark{a}\\ 
\colhead{} & \colhead{(J2000)} & \colhead{(J2000)} & \colhead{(mag)} & \colhead{(mag)} & \colhead{(mag)} & \colhead{} & \colhead{} & \colhead{} & \colhead{(sec)} & \colhead{(mag/s/\arcsec$^2$)} &\colhead{} & \colhead{(sec)} & \colhead{(mag s$^{-1}$ \arcsec$^2$)} \\
\colhead{(1)} & \colhead{(2)} & \colhead{(3)} & \colhead{(4)} & \colhead{(5)} & \colhead{(6)} & \colhead{(7)} & \colhead{(8)} & \colhead{(9)} & \colhead{(10)} & \colhead{(11)} & \colhead{(12)} & \colhead{(13)} & \colhead{(14)} } 

\startdata
F2M1341+3301   &  13:41:08.11 &  +33:01:10.2   &  17.83 &  16.81 &  1.00 &  1.715 &  1 &  F105W &   897 & 23.8 & F160W & 1597 & 24.0 \\ 
F2M1359+3157   &  13:59:41.18 &  +31:57:40.5   &  17.86 &  16.72 &  0.88 &  1.724 &  1 &  F105W &   897 & 23.6 & F160W & 1597 & 23.9 \\ 
F2M1036+2828   &  10:36:33.54 &  +28:28:21.6   &  17.93 &  17.15 &  0.85 &  1.762 &  1 &  F105W &   897 & 23.6 & F160W & 1597 & 24.0 \\ 
F2M1344+2839   &  13:44:08.31 &  +28:39:32.0   &  17.48 &  16.66 &  0.30 &  1.770 &  1 &  F105W &   897 & 24.0 & F160W & 1597 & 24.0 \\ 
F2M0921+1918   &  09:21:45.69 &  +19:18:12.6   &  17.70 &  16.48 &  1.20 &  1.791 &  1 &  F105W &   897 & 23.3 & F160W & 1597 & 23.4 \\ 
F2M0738+2750   &  07:38:20.10 &  +27:50:45.5   &  17.99 &  17.18 &  0.99 &  1.985 &  1 &  F105W &   897 & 23.9 & F160W & 1597 & 24.0 \\ 
F2M1427+3723   &  14:27:44.34 &  +37:23:37.5   &  18.09 &  16.99 &  0.60 &  2.168 &  1 &  F105W &   997 & 23.9 & F160W & 1597 & 23.9 \\ 
F2M0943+5417   &  09:43:17.68 &  +54:17:05.5   &  16.97 &  16.15 &  0.30 &  2.232 &  2 &  F105W &  2194 & 24.2 & F160W & 3193 & 24.2 \\ 
F2M2222$-$0202 &  22:22:52.79 &  $-$02:02:57.4 &  18.39 &  17.07 &  0.77 &  2.252 &  2 &  F125W &  1794 & 24.2 & F160M & 3194 & 24.3 \\ 
F2M1531+2423   &  15:31:50.47 &  +24:23:17.6   &  17.58 &  16.60 &  0.75 &  2.287 &  2 &  F125W &  1794 & 24.2 & F160W & 3194 & 24.3 \\ 
UKFS0030+0025  &  00:30:04.96 &  +00:25:01.3   &  19.32 &  18.01 &  0.90 &  2.299 &  2 &  F125W &  1794 & 23.9 & F160W & 2794 & 24.0 \\ 
\enddata

\tablenotetext{a}{The 3$\sigma$ surface brightness limit quoted here is a per pixel value ($0.06\arcsec \times 0.06\arcsec$)}


\end{deluxetable*}

\begin{deluxetable*}{ccccc}
\tablecaption{Point Spread Function Parameters \label{tab:psf}}
\tablewidth{0pt} 
\tablehead{\colhead{Filter} & \multicolumn{2}{c}{Number of Fields} & \colhead{$N_{\rm stars}$} & \colhead{FWHM} \\ 
\colhead{} & \colhead{Archive} & \colhead{Program} & \colhead{} & \colhead{(\arcsec)} } 
\startdata
F105W & 7 & 8 & 35 & 0.147 \\
F125W & 9 & 3 & 29 & 0.2094 \\
F160W & 9 & 11 & 46 & 0.1782 \\
\enddata
\end{deluxetable*}




\begin{deluxetable}{cccccccccccc}


\tabletypesize{\small}

\tablewidth{0pt} 

\tablecaption{Galfit Multi-Component Fitting Parameters  \label{tab:galfit_params}}


\tablehead{\colhead{Source} & \colhead{Model} & \colhead{$Y$ } & \colhead{$R_e$\tablenotemark{a}} & \colhead{$n$} & \colhead{Dist} & \colhead{$\chi^2_{\rm red}$} & \colhead{$H$} & \colhead{$R_e$\tablenotemark{a}} & \colhead{$n$} & \colhead{Dist} & \colhead{$\chi^2_{\rm red}$} \\ 
\colhead{Name} & \colhead{Type} & \colhead{(mag)} & \colhead{(pixels)} & \colhead{} & \colhead{(\arcsec)} & \colhead{} & \colhead{(mag)} & \colhead{(pixels)} & \colhead{} & \colhead{(\arcsec)} & \colhead{} } 

\startdata
F2M1341  &PSF\tablenotemark{c} & 18.59$\pm$0.25 &   \ldots       &  \ldots                &   0    &   4.777 & 17.3$\pm$0.01  &   \ldots       &   \ldots              &   0    &  5.430 \\
Comp     &   \sersic           & 23.24$\pm$0.13 & 12.7$\pm$2.3   & 1.31$\pm$0.26          &   1.35 &  \ldots & 22.3$\pm$1.2   & 61.1$\pm$144.8 & 7.64$\pm$6.10         &   1.44 & \ldots \\
Comp     &   \sersic           & 21.69$\pm$0.57 & 79.7$\pm$69.5  & 4.8$\pm$1.6            &   1.59 &  \ldots & 22.4$\pm$0.4   & 12.6$\pm$0.7   & 0.64$\pm$0.08         &   2.06 & \ldots \\
F2M1359  &PSF\tablenotemark{b} & 20.74$\pm$0.31 &   \ldots       &  \ldots                &   0    &   2.407 & 18.48$\pm$0.31 &   \ldots       &   \ldots              &   0    & 13.505 \\
Host     &   \sersic           & 21.51$\pm$0.31 & 3.5$\pm$5.8    & 2.54$\pm$0.40          &   0.11 &  \ldots & 19.6$\pm$1.4   & 425$\pm$1015   & 8.0$\pm$4.9           &   0.19 & \ldots \\
F2M1036  &PSF\tablenotemark{b} & 18.58$\pm$0.01 &   \ldots       &  \ldots                &   0    &   2.917 & 17.97$\pm$0.01 &   \ldots       &   \ldots              &   0    &  6.003 \\
Comp     &   \sersic           & 22.23$\pm$0.01 & 5.55$\pm$0.06  & 0.30$\pm$0.03          &   1.46 &  \ldots & 21.24$\pm$0.01 & 8.20$\pm$0.06  & 0.20$\pm$0.02         &   1.38 & \ldots \\
Comp     &   \sersic           & 21.84$\pm$0.01 & 7.22$\pm$0.07  & 0.13$\pm$0.01          &   0.81 &  \ldots & 22.75$\pm$0.06 & 11.0$\pm$0.7   & 0.92$\pm$0.14         &   0.77 & \ldots \\
F2M1344  &PSF\tablenotemark{b} & 18.42$\pm$0.26 &   \ldots       &   \ldots               &   0    &   5.652 & 18.39$\pm$0.16 &   \ldots       &   \ldots              &   0    & 11.061 \\
Host     &   \sersic           & 22.14$\pm$0.39 & 9.2$\pm$6.0    &  1.03$\pm$0.71         &   0.54 &  \ldots & 19.25$\pm$0.55 & 4$\pm$220      & 0.01\tablenotemark{d} &   0.02 & \ldots \\
Comp     &   \sersic           & 19.6$\pm$1.6   & 3202$\pm$12685 & 20.00\tablenotemark{d} &   1.21 &  \ldots & 20.13$\pm$0.67 & 233$\pm$269    & 7.17$\pm$2.25         &   1.25 & \ldots \\
Comp     &   \sersic           & 22.85$\pm$0.04 & 6.64$\pm$0.16  & 0.10$\pm$0.05          &   1.58 &  \ldots & 22.15$\pm$0.02 & 7.7$\pm$5.1    & 0.04\tablenotemark{d} &   1.52 & \ldots \\
F2M0921  &PSF\tablenotemark{b} & 19.13$\pm$0.16 &   \ldots       &   \ldots               &   0    &   3.905 & 17.54$\pm$0.16 &   \ldots       &   \ldots              &   0    &  9.269 \\
Host     &   \sersic           & 20.16$\pm$0.31 & 4$\pm$62       & 0.01\tablenotemark{d}  &   0.06 &  \ldots & 19.47$\pm$0.32 & 6.1$\pm$2.3    & 19.3\tablenotemark{d} &   0.31 & \ldots \\
Comp     &   \sersic           & 21.98$\pm$0.07 & 10.1$\pm$1.1   & 2.43$\pm$0.26          &   1.23 &  \ldots & 19.79$\pm$0.67 & 193$\pm$271    & 10.9\tablenotemark{d} &   1.45 & \ldots \\
Comp     &   \sersic           & 18.69$\pm$1.84 & 3570$\pm$11295 & 11.74\tablenotemark{d} &   2.10 &  \ldots & 20.1$\pm$1.5   & 408$\pm$1325   & 12.3\tablenotemark{d} &   2.33 & \ldots \\
AGN?     &   PSF               & 21.27$\pm$0.01 &   \ldots       &   \ldots               &   4.13 &  \ldots & 21.38$\pm$0.18 &   \ldots       &   \ldots              &   4.11 & \ldots \\
Comp     &   \sersic           &  \ldots        &   \ldots       &   \ldots               & \ldots &  \ldots & 21.12$\pm$0.29 & 23$\pm$26      & 11.7\tablenotemark{d} &   4.15 & \ldots \\
F2M0738  &PSF\tablenotemark{b} & 18.63$\pm$0.26 &   \ldots       &   \ldots               &   0    &   3.287 & 17.50$\pm$0.27 &   \ldots       &   \ldots              &   0    &  3.047 \\
Host     &   \sersic           & 22.22$\pm$0.39 & 16.9$\pm$6.3   & 1.85$\pm$0.73          &   0.55 &  \ldots & 22.92$\pm$0.80 & 3$\pm$6        & 2.8$\pm$1.4           &   0.46 & \ldots \\
Comp     &   \sersic           & 21.36$\pm$0.02 & 7.02$\pm$0.17  & 2.13$\pm$0.06          &   1.86 &  \ldots & 20.65$\pm$0.01 & 6.60$\pm$0.14  & 3.93$\pm$0.11         &   1.84 & \ldots \\
F2M1427  &   PSF               & 18.56$\pm$0.01 &   \ldots       &   \ldots               &   0    &   6.084 & 17.70$\pm$0.01 &   \ldots       &   \ldots              &   0    &  5.220 \\
Comp     &   \sersic           & 21.82$\pm$0.08 & 15.25$\pm$2.17 & 3.68$\pm$0.38          &   2.26 &  \ldots & 19.95$\pm$0.43 & 287$\pm$274    & 11.86\tablenotemark{d}&   2.27 & \ldots \\
Comp     &   \sersic           & 24.36$\pm$0.08 & 1.91$\pm$0.34  & 1.79$\pm$0.84          &   1.38 &  \ldots & 23.61$\pm$0.11 & 2.1$\pm$0.8    & 6$\pm$3               &   1.39 & \ldots \\
Comp     &   \sersic           & 23.11$\pm$0.12 & 9.7$\pm$1.7    & 1.78$\pm$0.33          &   1.54 &  \ldots & 21.22$\pm$0.23 & 40$\pm$15      & 4.6$\pm$0.8           &   1.60 & \ldots \\
F2M0943  &PSF\tablenotemark{c} & 17.57$\pm$0.31 &   \ldots       &   \ldots               &   0    &  15.564 & 16.85$\pm$0.01 &   \ldots       &   \ldots              &   0    & 23.063 \\
F2M2222  &   PSF               & 18.40$\pm$0.01 &   \ldots       &   \ldots               &   0    &   2.889 & 17.97$\pm$0.01 &   \ldots       &   \ldots              &   0    &  9.699 \\
Comp     &   \sersic           & 21.56$\pm$0.02 & 12.32$\pm$0.34 & 1.62$\pm$0.05          &   3.93 &  \ldots & 21.53$\pm$0.03 & 11.65$\pm$0.52 & 1.64$\pm$0.08         &   3.93 & \ldots \\
F2M1531  &PSF\tablenotemark{b} & 18.06$\pm$0.16 &   \ldots       &   \ldots               &   0    &   3.912 & 17.75$\pm$0.30 &   \ldots       &   \ldots              &   0    &  7.646 \\
Host     &   \sersic           & 20.01$\pm$0.15 & 4.4$\pm$5.2    & 3.64$\pm$0.39          &   0.19 &  \ldots & 18.86$\pm$0.31 & 1.9$\pm$5.9    &  4.48$\pm$0.33        &   0.12 & \ldots \\
Comp     &   \sersic           & 24.17$\pm$0.04 & 5.09$\pm$0.33  & 0.6$\pm$0.2            &   1.55 &  \ldots & 23.93$\pm$0.04 & 5.4$\pm$0.4    &  0.61$\pm$0.25        &   1.54 & \ldots \\
UKFS0030 &     PSF             & 19.55$\pm$0.31 &   \ldots       &   \ldots               &   0    &   1.997 & 18.72$\pm$0.01 &   \ldots       &   \ldots              &   0    &  4.238 \\
Host     &  \sersic            & 21.11$\pm$0.30 & 2.9$\pm$5.9    & 2.74$\pm$0.36          &   0.11 &  \ldots &  \ldots        &   \ldots       &   \ldots              & \ldots & \ldots \\
Comp     &  \sersic            & 22.52$\pm$0.01 & 5.48$\pm$0.11  & 0.65$\pm$0.05          &   1.35 &  \ldots & 22.38$\pm$0.01 & 5.83$\pm$0.13 &  0.71$\pm$0.06         &   1.30 & \ldots \\
\enddata
\tablenotetext{a}{The distance to the \sersic\ radius in pixels, which can be converted to arcseconds using the image scale of 0.060\arcsec/pixel.}
\tablenotetext{b}{The PSF magnitude reprted here is comprised of the integrated flux from two PSF components, converted to a magnitude using equation \ref{eqn:phot_2psf}.}
\tablenotetext{c}{The PSF magnitude reprted here is comprised of the integrated flux from PSF component plus a \sersic\ component that is sharply concentrated at the position of the PSF, with unphysical parameters.  The total PSF magnitude is computed with equation \ref{eqn:phot_psf}.}
\tablenotetext{d}{These parameters are flagged by Galfit as being outside the range of acceptable values, however, the fit resulted in an acceptable $\chi^2_{\rm red}$ enabling a capture of the total flux in the host components. We do not report errors for these parameters.}



\end{deluxetable}




\begin{deluxetable}{ccccccccccc}



\tablewidth{0pt}

\tablecaption{Quasar Magnitudes and Colors \label{tab:psf_mags}}


\tablehead{\colhead{} & \multicolumn{3}{c}{SExtractor {\tt MAG\_AUTO}} & \multicolumn{3}{c}{GALFIT primary PSF\tablenotemark{a}} & \multicolumn{3}{c}{GALFIT combined PSF\tablenotemark{b}} & \colhead{}\\
\colhead{Name} & \colhead{$U$} & \colhead{$V$} & \colhead{$U-V$} & \colhead{$U$} & \colhead{$V$} & \colhead{$U-V$} & \colhead{$U$} & \colhead{$V$} & \colhead{$U-V$} & \colhead{$E(B-V)$} \\ 
\colhead{} & \colhead{(mag)} & \colhead{(mag)} & \colhead{(mag)} & \colhead{(mag)} & \colhead{(mag)} & \colhead{(mag)} & \colhead{(mag)} & \colhead{(mag)} & \colhead{(mag)} & \colhead{(mag)} } 

\startdata
  F2M1341  & 18.66$\pm$0.07 & 17.38$\pm$0.02 & 1.28$\pm$0.02 & 19.24$\pm$0.25 & 17.29$\pm$0.01 & 1.95$\pm$0.04 & 18.59$\pm$0.25 & 17.29$\pm$0.01 & 1.30$\pm$0.06 &  0.57 \\ 
  F2M1359 & 20.51$\pm$0.06 & 18.27$\pm$0.03 & 2.24$\pm$0.03 & 20.99$\pm$0.31 & 18.50$\pm$0.31 & 2.49$\pm$0.28 & 20.74$\pm$0.31 & 18.48$\pm$0.31 & 2.26$\pm$0.28 &  0.50 \\ 
  F2M1036  & 18.67$\pm$0.03 & 18.11$\pm$0.02 & 0.55$\pm$0.02 & 19.12$\pm$0.01 & 18.51$\pm$0.01 & 0.61$\pm$0.01 & 18.58$\pm$0.01 & 17.97$\pm$0.01 & 0.61$\pm$0.01 &  0.47 \\ 
  F2M1344  & 18.41$\pm$0.03 & 17.78$\pm$0.02 & 0.63$\pm$0.02 & 19.02$\pm$0.26 & 18.50$\pm$0.16 & 0.52$\pm$0.14 & 18.42$\pm$0.26 & 18.39$\pm$0.16 & 0.03$\pm$0.15 &  0.07 \\ 
  F2M0921 & 19.01$\pm$0.03 & 17.62$\pm$0.02 & 1.40$\pm$0.02 & 19.34$\pm$0.16 & 17.60$\pm$0.16 & 1.74$\pm$0.14 & 19.13$\pm$0.16 & 17.54$\pm$0.16 & 1.59$\pm$0.13 &  0.65 \\
  F2M0738  & 18.64$\pm$0.02 & 17.63$\pm$0.02 & 1.01$\pm$0.02 & 19.25$\pm$0.26 & 17.68$\pm$0.27 & 1.57$\pm$0.22 & 18.63$\pm$0.26 & 17.50$\pm$0.27 & 1.13$\pm$0.21 &  0.49 \\
  F2M1427  & 18.58$\pm$0.02 & 17.85$\pm$0.02 & 0.73$\pm$0.01 & 18.56$\pm$0.01 & 17.70$\pm$0.01 & 0.86$\pm$0.01 & 18.56$\pm$0.01 & 17.70$\pm$0.01 & 0.86$\pm$0.01 &  0.27 \\ 
  F2M0943  & 17.70$\pm$0.02 & 16.91$\pm$0.07 & 0.78$\pm$0.05 & 18.75$\pm$0.31 & 18.10$\pm$0.01 & 0.65$\pm$0.11 & 17.57$\pm$0.31 & 16.85$\pm$0.01 & 0.72$\pm$0.10 &  0.11 \\ 
F2M2222  & 18.51$\pm$0.02 & 17.99$\pm$0.02 & 0.52$\pm$0.01 & 18.40$\pm$0.01 & 17.97$\pm$0.01 & 0.43$\pm$0.01 & 18.40$\pm$0.01 & 17.97$\pm$0.01 & 0.43$\pm$0.01 &  0.33 \\ 
  F2M1531  & 18.00$\pm$0.02 & 17.50$\pm$0.02 & 0.51$\pm$0.01 & 18.68$\pm$0.16 & 17.96$\pm$0.30 & 0.72$\pm$0.20 & 18.06$\pm$0.16 & 17.75$\pm$0.30 & 0.30$\pm$0.18 &  0.32 \\
UKFS0030 & 19.71$\pm$0.09 & 18.87$\pm$0.03 & 0.84$\pm$0.04 & 19.55$\pm$0.31 & 18.72$\pm$0.01 & 0.83$\pm$0.10 & 19.55$\pm$0.31 & 18.72$\pm$0.01 & 0.83$\pm$0.10 &  0.32 \\ 
\enddata

\tablenotetext{a}{In sources where two PSFs were used to fit the quasar flux, we consider the brighter fitted component to be the primary component.}
\tablenotetext{b}{These magnitudes include integrated quasar magnitudes arising from fitting a single PSF, two PSFs, or a PSF  plus \sersic\ components.}
\tablecomments{The magnitudes presented in this table are directly measured from the {\em HST} blue (F105W or F125W) and red (F160W) bandpasses as pseudo rest-frame $U$ and $V$ magnitudes, respectively. } 

\end{deluxetable}




\begin{deluxetable*}{cccccccc}



\tablewidth{0pt}

\tablecaption{Non-Parametric Measures for Red Quasar Host Galaxies \label{tab:nonparam}}


\tablehead{\colhead{Object} & \multicolumn{3}{c}{Blue Filter}  &  \multicolumn{3}{c}{Red Filter}  & \colhead{} \\ 
\colhead{Name} & \colhead{$R_{\rm Pet}$} & \colhead{Gini} & \colhead{$M_{20}$} & \colhead{$R_{\rm Pet}$} & \colhead{Gini} & \colhead{$M_{20}$} & \colhead{$E(B-V)$} \\ 
\colhead{} & \colhead{(arcsec)} & \colhead{} & \colhead{} & \colhead{(arcsec)} & \colhead{} & \colhead{} & \colhead{(mag)} } 

\startdata
F2M1341 & 1.55 & 0.61 & $-$0.39 & 0.44 & 0.43 & $-$0.39 & 0.57 \\
F2M1359 & 2.12 & 0.48 & $-$0.60 & 2.12 & 0.49 & $-$0.61 & 0.50 \\
F2M1036 & 0.41 & 0.22 & $-$0.73 & 0.43 & 0.25 & $-$0.78 & 0.47 \\
F2M1344 & 0.36 & 0.37 & $-$0.47 & 0.51 & 0.57 & $-$0.55 & 0.07 \\
F2M0921 & 1.76 & 0.61 & $-$0.38 & 2.05 & 0.62 & $-$0.35 & 0.65 \\
F2M0738 & 2.11 & 0.64 & $-$0.66 & 0.43 & 0.31 & $-$0.76 & 0.49 \\
F2M1427 & 0.43 & 0.50 & $-$0.47 & 0.42 & 0.20 & $-$0.60 & 0.27 \\
F2M0943 & 0.42 & 0.20 & $-$0.52 & 0.51 & 0.59 & $-$0.27 & 0.11 \\
F2M2222 & 0.41 & 0.52 & $-$0.33 & 0.53 & 0.54 & $-$0.26 & 0.33 \\
F2M1531 & 1.16 & 0.71 & $-$0.26 & 0.92 & 0.60 & $-$0.33 & 0.32 \\
UKFS0030 & 1.66 & 0.66 & $-$0.40 & 0.42 & 0.39 & $-$0.84 & 0.32 \\
\enddata




\end{deluxetable*}


\clearpage

\end{document}